\documentclass[11pt]{article}
\usepackage[utf8]{inputenc}

\usepackage{fullpage}
\usepackage[margin = 2.5cm]{geometry}
\usepackage{tcolorbox}

\usepackage{amsmath, amssymb, amsthm}
\usepackage{mathtools}  
\usepackage{xfrac} 

\usepackage{graphicx}
\usepackage{color}
\usepackage{xcolor}
\definecolor{Blue}{HTML}{2D2F92}

\usepackage[hyperindex,breaklinks]{hyperref}
\hypersetup{
    colorlinks = true,
    linkcolor = black,
    citecolor = Blue,
    urlcolor = black
}
\providecommand{\email}[1]{\href{mailto:#1}{\nolinkurl{#1}}}

\usepackage{nicefrac}
\usepackage{subfig}

\newtheorem{theorem}{Theorem}[section]
\newtheorem{claim}[theorem]{Claim}
\newtheorem{corollary}[theorem]{Corollary}
\newtheorem{proposition}[theorem]{Proposition}
\newtheorem{lemma}[theorem]{Lemma}
\newtheorem{definition}[theorem]{Definition}

\DeclareMathOperator{\E}{\mathbf{E}}
\DeclareMathOperator{\Var}{\mathbf{Var}}

\newcommand{\given}{\mid}
\DeclareMathOperator{\Ball}{Ball}
\DeclareMathOperator{\tr}{tr}

\newcommand {\roundup}   [1] {{\lceil {#1} \rceil}}
\newcommand {\rounddown} [1] {{\lfloor {#1} \rfloor}}

\newcommand{\nL}{\tilde L}     
\newcommand{\Lap}{L}  

\newcommand{\bbR}{\mathbb{R}}

\newcommand{\calD}{\mathcal{D}}
\newcommand{\calE}{\mathcal{E}}

\newcommand{\calI}{\mathcal{I}}

\newcommand{\calN}{\mathcal{N}}

\newcommand {\ONE}      {\text{\textbf{1}}}

\newcommand{\relbecause}[2]{\stackrel{\text{\tiny
#2}}{#1}}
\newcommand{\eqbecause}[1]{\relbecause{=}{#1}}

\newcommand {\FFF} {\bar\Phi}
\newcommand {\nc} {\frac{1}{\sqrt{2\pi}}}
\newcommand {\ncc} [2] {\frac{#1}{\sqrt{2\pi} #2}}

\title{Higher-Order Cheeger  Inequality for Partitioning with Buffers}
 \author{
 Konstantin Makarychev\thanks{The author was supported by the NSF Awards CCF-1955351, CCF-1934931, and EECS-2216970.} \\Northwestern
\and Yury Makarychev\thanks{The author was supported by the NSF Awards  
CCF-1955173, CCF-1934843, and ECCS-2216899.
}\\TTIC
\and Liren  Shan\thanks{The author was supported by the NSF Awards CCF-1955351, CCF-1934931, and EECS-2216970.}\\Northwestern
\and Aravindan Vijayaraghavan\thanks{The author was supported by the NSF Awards CCF-1934931, CCF-2154100, and EECS-2216970.}\\Northwestern
}
\date{}
\begin{document}
\maketitle
\thispagestyle{empty}
\begin{abstract}


We prove a new generalization of the higher-order Cheeger inequality for partitioning with buffers. 
Consider a graph $G=(V,E)$.
The buffered expansion of a set $S \subseteq V$ with a buffer $B \subseteq V \setminus S$ is the edge expansion of $S$ after removing all the edges from set $S$ to its buffer $B$. 
An $\varepsilon$-buffered $k$-partitioning is a partitioning of a graph into disjoint components $P_i$ and buffers $B_i$, in which the size of buffer $B_i$ for $P_i$ is small relative to the size of $P_i$: 
$|B_i| \le \varepsilon |P_i|$.
The buffered expansion of a buffered partition is the maximum of buffered expansions of the $k$ sets $P_i$ with buffers $B_i$.
Let $h^{k,\varepsilon}_G$ be the buffered expansion of the optimal $\varepsilon$-buffered $k$-partitioning, then for every $\delta>0$,  $$h_G^{k,\varepsilon} \le  O_\delta(1) \cdot \Big( \frac{\log k}{ \varepsilon}\Big) \cdot \lambda_{\lfloor (1+\delta) k\rfloor},$$  
where $\lambda_{\lfloor (1+\delta)k\rfloor}$ is the $\lfloor (1+\delta)k\rfloor$-th smallest eigenvalue of the normalized Laplacian of $G$.

Our inequality is constructive and avoids the ``square-root loss'' that is present in the standard Cheeger inequalities (even for $k=2$). We also provide a complementary lower bound,
and a novel generalization to the setting with arbitrary vertex weights and edge costs. 
Moreover our result implies and generalizes the standard higher-order Cheeger inequalities and another recent Cheeger-type inequality by Kwok, Lau, and Lee (2017) involving robust vertex expansion.

\end{abstract}
\newpage
\newpage
\setcounter{page}{1}

\section{Introduction}
Cheeger's inequality is a fundamental result in spectral graph theory that relates the connectivity of a graph to the eigenvalues of the Laplacian matrix associated with the graph. Consider an undirected $d$-regular graph $G=(V,E)$ on $n$ vertices. Let $L_G$ be the normalized
Laplacian of the graph defined by $L_G = I - \tfrac{1}{d} A$, where $A$ is the adjacency matrix of the graph $G$. Let $0=\lambda_1 \le  \lambda_2 \le \lambda_3 \dots \le \lambda_n \le 2$ be the eigenvalues of $L_G$. For every vector $z \in {\mathbb R}^V$ with coordinates $z(u)$ (where $u\in V$),
\begin{equation}\label{eq:lapl}
    z^TL_Gz = \frac{1}{d}\sum_{(u,v)\in E} (z(u) - z(v))^2.
\end{equation}
For a set $S \subseteq V$, let $\delta_G(S,V \setminus S)$ denote the number of edges in the graph crossing the cut $(S, V\setminus S)$. The Cheeger constant or expansion of the graph $G$ is
$$ h_G \coloneqq \min_{S \subseteq V: |S|\le |V|/2} \phi_G(S), \text{ where } \phi_G(S) \coloneqq  \frac{\delta_G(S, V \setminus S)}{d|S|}, $$
is called the expansion of the cut $S, V\setminus S$.   
Cheeger's inequality by Alon and Milman~\cite{AlonM85, Alon86, Cheeger} states that 
\begin{equation}\label{eq:basic:cheeger}
    \frac{\lambda_2}{2} \le h_G \le \sqrt{2 \lambda_2}.
\end{equation}
Similar inequalities also hold for graph partitioning into $k$ parts \cite{LouisRTV12, LeeOvT12}. Here is a
\emph{higher order} Cheeger inequality by Lee, Oveis-Gharan and Trevisan~\cite{LeeOvT12} (see also the paper~\cite{LouisRTV12} by Louis, Raghavendra, Tetali and Vempala): 
For every $\delta>0$,
\footnote{The upper bound on $h_G^k$ in \cite{LouisRTV12} is $O(\sqrt{\log k})\sqrt{\lambda_{ck}}$ where $c> 1$ is an absolute constant.} 
\begin{equation}\label{eq:basic:higherorder}
   \frac{\lambda_k}{2} \le h_G^k \le O_\delta\big(\sqrt{\log k}\big) 
   \cdot \sqrt{\lambda_{\rounddown{(1+\delta)k}}},
\end{equation}
where 
$\lambda_i$ is the $i$-th smallest eigenvalue of the normalized Laplacian $L_G$, and
$$ h_G^k = \min_{\substack{\text{partitions } \\ P_1, \dots, P_k  \text{ of } V}} ~~\max_{i \in [k]} \phi_G(P_i).$$
The upper bounds in 
\eqref{eq:basic:cheeger}
and \eqref{eq:basic:higherorder} are constructive, which means that there is a polynomial-time algorithm that finds a partitioning $P_1,\dots, P_k$ using a spectral embedding of $G$, an embedding of the graph vertices into $\mathbb{R}^{k'}$ based on the first $k'=\rounddown{(1+\delta)}k$ eigenvectors of the Laplacian. Similar spectral algorithms are commonly used in practice~\cite{NgJW01,mcsherry2001}.
We refer the reader to examples of applications of Cheeger's inequality to spectral clustering~\cite{KannanVV04, SpielmanT07, Spielman07}, image segmentation~\cite{ShiM00}, random sampling and approximate counting~\cite{SinclairJ89}.
Cheeger's inequality is widely used in combinatorics and graph theory. Higher-order Cheeger inequalities also have connections to the small-set expansion conjecture~\cite{RaghavendraS10, RaghavendraST12}, an important problem in the area of approximation algorithms.

The objective of abovementioned $k$-way graph partitioning algorithms is to find  the Sparsest $k$-Partition of the graph i.e., a partition $P_1, \dots, P_k$ that {\em minimizes} the value of $\max_{i \in [k]} \phi_G(P_i)$. Together the lower and upper bounds \eqref{eq:basic:higherorder} give a bound on the cost of the algorithmic solution in terms of the optimal solution: 
$\max_{i \in [k]} \phi_G(P_i)\leq O_\delta\Big(\sqrt{\log k \cdot  h_{G}^{\rounddown{(1+\delta)k}}}\Big)$.  This bound may be good for large values of 
$h_G^{\rounddown{(1+\delta)k}}$ but can also be really bad for small values of 
$h_G^{\rounddown{(1+\delta)k}}$. In fact, the approximation factor of such $k$-way partitioning algorithm may be as large as $\Omega(n)$ even for $k=2$. It  can be so large because the upper bound is non-linear -- it has a ``square-root loss''. To address this problem, several improved Cheeger inequalities under additional structural assumptions on the graph $G$ have been presented in the literature~\cite{KwokLLOT13, KwokLL17}. 

In this work, we introduce a new type of graph partitioning -- partitioning with buffers -- and prove a higher-order Cheeger inequality for them. Our inequality avoids the ``square-root loss'' and provides a constant bi-criteria approximation algorithm for the problems (see below for details). 
While being a natural problem, in and of itself, our results for buffered partitioning also
imply 
the standard higher-order Cheeger inequality (\ref{eq:basic:higherorder}) and a Cheeger-type inequality by Kwok, Lau, and Lee~\cite{KwokLL17} for robust vertex expansion (see Section~\ref{sec:intro-robust}). Finally, these Cheeger inequalities can also be extended to a more general setting with arbitrary vertex weights and edge costs: in contrast, we are not aware of such a generalization for the standard Cheeger inequalities i.e., without buffers. 

\subsection{Cheeger inequality for Buffered Partitions}
To simplify the exposition, we first present and discuss the setting where $G$ is a $d$-regular graph. Then, in Section~\ref{sec:intro:generalweights}, we consider non-regular graphs $G$ with arbitrary positive vertex weights and edge costs.

\paragraph{Multi-way Partitioning with Buffers.}
For every $\varepsilon \in [0,1)$, an {\em $\varepsilon$-buffered $k$-partitioning} of an undirected graph $G=(V,E)$ is a collection of subsets $P_1, P_2, \dots, P_k\subset V$ and $B_1, B_2, \dots, B_k \subset V$ that satisfy the following conditions:
\begin{enumerate}
    \item All sets $P_i$ and $B_j$ are pairwise disjoint (i.e., $P_i\cap P_j =\varnothing$, $B_i\cap B_j=\varnothing$, and $P_i\cap B_j=\varnothing$ for all $i,j\in\{1,\dots,k\}$);
    \item $\bigcup_{i=1}^k(P_i \cup B_i) = V$;
    \item Sets $P_i$ are nonempty;
    \item $|B_{i}|\leq \varepsilon |P_i|$ for all $i\in\{1,\dots,k\}$.
\end{enumerate}

\noindent 
We say that $B_i$ is \emph{the buffer} for $P_i$. We denote this buffered partition by $(P_1, \dots, P_k \parallel B_1, \dots, B_k)$. Now we define the buffered expansion of a set $P$ with buffer $B$ for $d$-regular graphs. Later, we will extend this definition to graphs with \emph{arbitrary vertex weights and edge costs}. The buffered expansion of a set $P$ with buffer $B$
$$
\phi_G(P \parallel B) = \frac{\delta_G\big(P, V \setminus (P \cup B)\big)}{d |P|}.
$$
The definition is similar to that of the standard set expansion except we do not count edges from set $S$ to its buffer $B$.
Define the cost $\phi_G(P_1, \dots, P_k \parallel B_1, \dots, B_k)$ of a buffered partition:
\begin{equation} \label{eq:buffered:expansion}
\phi_G(P_1, \dots, P_k \parallel B_1, \dots, B_k) = \max_{i \in \{1, \dots, k\}} \phi_G(P_i \parallel B_i).
\end{equation}

See Figure~\ref{fig:buffer} on page \pageref{fig:buffer} for an illustration of the edges that contribute towards the expansion $\phi_G(P_i \parallel B_i)$. 
The $\varepsilon$-buffered expansion of the graph $G=(V,E)$ is defined as the minimum value among all $\varepsilon$-buffered partitions:
\begin{equation}\label{eq:h-G-k-eps}
 h_G^{k, \varepsilon} = \min_{\substack{\varepsilon\text{-buffered }k-\text{partition}\\ (P_1, \dots, P_k \parallel B_1, \dots, B_k)}} ~~\phi_G(P_1, \dots, P_k \parallel B_1, \dots, B_k).
\end{equation}

Our main result is a new Cheeger-type inequality that relates buffered expansion to the eigenvalues of the Laplacian. We first state it for regular graphs. Consider a $d$-regular graph $G$. Let $L_G$ be its normalized Laplacian and $0=\lambda_1 \leq \lambda_2 \leq \dots \leq \lambda_n$ be its eigenvalues.

\begin{theorem}\label{thm:main}
For every $\delta \in (0,1)$, 
\begin{equation}\label{eq:main:higherorder}
  h_G^{k,\varepsilon} \leq \frac{c(\delta) \log k}{\varepsilon} 
   \cdot \lambda_{\lfloor(1+\delta)k\rfloor},
\end{equation}
where $c(\delta)$ is a function that depends only on $\delta$.
Furthermore, there is a randomized polynomial-time algorithm that given $G$ finds an $\varepsilon$-buffered $k$-partitioning $(P_1, \dots, P_k\parallel B_1, \dots, B_k)$ with $\phi_G(P_1, \dots, P_k \parallel B_1, \dots, B_k) \le \frac{c(\delta) \log k}{\varepsilon}  \lambda_{\lfloor(1+\delta)k\rfloor}$. 
\end{theorem}
As in the standard Cheeger-type inequality (\ref{eq:basic:higherorder}), we upper bound expansion for $k$-way partitioning in terms of $\lambda_{k'}$, where $k' = \lfloor(1+\delta)k\rfloor$ may be larger than $k$ (depending on the value of $\delta > 0$). However, for every fixed $k$, we can let $\delta = 1/(k+1)$ and get the following result.    
\begin{corollary}\label{cor:main-no-loss-in-k}
For every $k$, 
$
  h_G^{k,\varepsilon} \leq \frac{c_k}{\varepsilon}  
   \cdot \lambda_{k}
$,
where $c_k$ depends only on $k$.
Furthermore, there is a randomized polynomial-time algorithm that given $G$ finds an $\varepsilon$-buffered $k$-partitioning\\ $(P_1, \dots, P_k \parallel B_1, \dots, B_k)$ with $\phi_G(P_1, \dots, P_k \parallel B_1, \dots, B_k) \le \frac{c_k}{\varepsilon}  \lambda_{k}$. 
\end{corollary}
 Theorem~\ref{thm:main-weighted} presented later is a novel generalization of Theorem~\ref{thm:main} to graphs with vertex weights and edge costs.

\paragraph{Approximation results}
The spectral graph partitioning algorithm provided by Theorem~\ref{thm:main} can be seen as an $O_{\delta}\Big(\dfrac{1}{\varepsilon}\log k\Big)$-\emph{pseudo-approximation} algorithm for the $k$-way sparsest partitioning problem. It finds an $\varepsilon$-buffered $k$-partitioning $(P_1,\dots,P_k,B_1,\dots,B_k)$ with the maximum expansion bounded by $O_{\varepsilon,\delta}(\log k)$ times the cost of the true optimum solution of the non-buffered $\rounddown{(1+\delta)k}$-way partitioning problem. That is, the solution produced by our algorithm has an approximation factor of $O_{\varepsilon,\delta}(\log k)$ but (1) uses $\varepsilon$ buffers around each set $P_i$, and (2) has fewer sets than the true optimal solution. This pseudo-approximation algorithm also works for non-regular graphs with vertex weights and edge costs. See Theorem~\ref{thm:approx-sparsest-k-partitioning} for details.
Applying this pseudo-approximation algorithm recursively, we get an $O(1/\varepsilon)$-pseudo-approximation algorithm for the Buffered Balanced Cut problem (see Theorem~\ref{thm:buffer-balanced}) and an $O(\log^2 k)$ pseudo-approximation algorithm for a buffered variant of the balanced $k$-partitioning problem (see Corollary~\ref{cor:approx-k-balanced}).

%

Let us examine some applications of buffered partitioning and our techniques.

\paragraph{Applications}
Spectral algorithms are widely used across several application domains because they are very fast and scalable in practice~\cite{Pothen,VonL}.  
For example, a standard off-the-shelf package finds the first $100$ eigenvectors of the Twitter graph~\cite{leskovec2012dataset} in less than half a minute. This graph has $81$ thousand nodes and $1.3$ million edges. In contrast, 
linear programming and semidefinite programming based methods do not scale well and cannot handle such large graphs at the present time. 
This motivates the design of spectral algorithms for graph partitioning with stronger guarantees.
Our work demonstrates that one can achieve very good theoretical guarantees for Buffered Sparsest $k$-Partitioning. 

As mentioned earlier, the algorithms we present in this paper give an $O_{\varepsilon,\delta}(\log k)$-pseudo-appro\-xi\-ma\-tion for the Buffered Sparsest $k$-Partitioning problem, and a $O(1/\varepsilon)$-pseudo-approximation for the Buffered Balanced Cut problem (see Section~\ref{sec:buf-balanced-cut}). For constant $\varepsilon$, this corresponds to a constant factor approximation with buffers. For comparison,
the best known approximation guarantees for Balanced Cut or Sparsest $k$-Cut without buffers incur logarithmic factors in the number of vertices $n$.\footnote{For Balanced Cut without buffers, the best true approximation factor is $O(\log n)$~\cite{balanced-racke-1,balanced-racke-2}, and the best pseudo-approximation is $O(\sqrt{\log n})$~\cite{ARV}.} Similarly, the best known approximation for Sparsest $k$-Partitioning is $O_{\delta}(\sqrt{\log n\log k})$~\cite{LM14}.
The caveat is, of course, that our algorithm produces an $\varepsilon$-buffered partitioning but we compare its cost with the cost of the optimal non-buffered partitioning. 

In applications of graph partitioning and clustering, relaxing the partitioning using buffers is often benign and even natural. Let us consider the following application of graph partitioning. Suppose we have a graph whose nodes represent user profiles in a social network (like the Twitter graph we mentioned earlier) and edges represent connections between them (friends, followers, etc). We would like to assign these profiles to two machines so that each machine is assigned about the same number of profiles and the number of separated connections is minimized. These are common requirement for graph processing systems. In other words, we need to solve the Balanced Cut problem for the given graph. If we run our algorithm on this graph, we will get two parts $S$, $T$ and buffer $B$. We can store $S$ and $T$ on the first and second machines, respectively, and replicate nodes in $B$ on both machines. This way we will separate only nodes located in $S$ and $T$. 
Partitioning with buffers can be useful to obtain better solutions for several other applications such as resource allocation and scheduling, where graph partitioning is used.   

Moreover, in applications like community detection, it is common for the communities to have small overlaps~\cite{YangLeskovec1,YangLeskovec2}. Vertices belonging to multiple communities may correspond to influential or well-connected nodes, that would disproportionately affect the cost in a disjoint partition. While there has been much recent interest in detecting overlapping communities, it is challenging to obtain algorithmic guarantees in the overlapping setting (see~\cite{SBMO,Orecchia22} for different formulations and results on this problem); in particular, there are very few theoretical results for spectral algorithms even in average-case models.  
An $\varepsilon$-buffered partitioning with sets $S, T$ and buffer $B$ can be viewed as two overlapping communities $S'=S \cup B$ and $T' = T \cup B$ with  small overlap $|S \cap T| \le \varepsilon \min\{|S|, |T|\}$. Hence $\varepsilon$-buffered partitions capture overlapping communities and allow us to reason about spectral methods even in the overlapping setting (see also footnote~\ref{foot:bicriteria}). 

Finally, buffered partitioning is an interesting problem in its own right, it gives a common, versatile generalization that captures important  results in spectral graph theory including higher-order Cheeger inequalities and robust vertex expansion as described in the next few sections.  

\subsection{Graphs with vertex weights and edge costs}\label{sec:intro:generalweights}
In the standard Cheeger inequality, the weight of every vertex must be equal to the total weight of edges incident on it. For instance, in $d$-regular graphs, the weights of all vertices are equal to $d$. Surprisingly, we can generalize our variant of Cheeger's inequality to vertex weighted graphs. We show that the Cheeger inequality for buffered partitions also holds when graph $G=(V,E, w, c)$ has vertex weights $w_u > 0$ and edge costs $c_{uv} > 0$. In that case, we define the non-normalized Laplacian $\nL_G$ for $G$ as follows. $\nL_G(u,u)$ is the total cost of all edges incident on $u$ and $\nL_G(u,v) = - c_{uv}$ for $(u,v)\in E$; all other entries are zero. Then, for any vector $z \in \bbR^n$, we have
\begin{equation}\label{eq:non-normalized-L}
z^T\nL_G z = \sum_{(u,v)\in E} c_{uv} (z(u) - z(v))^2.
\end{equation}
Further, we define the weight matrix $D_w$ as follows: $D_w(u,u) = w_u$ and $D_w(u,v) = 0$ if $u\neq v$ ($D_w$ is a diagonal matrix). Finally, we define the normalized Laplacian $\Lap_G = D_w^{-1/2} \nL_G D_w^{-1/2}$. Note that
$$z^T\Lap_G z = \sum_{(u,v) \in E} c_{uv} \left(\frac{z(u)}{w_u^{1/2}} - \frac{z(v)}{w_v^{1/2}}\right)^2.$$

Denote the weight of a set of vertices $A$ by $w(A) = \sum_{u\in A} w_u$. We extend the definitions of  
$\delta_G(A,B)$, $\phi_G(P\parallel B)$, $\phi_G(P_1,\dots, P_k \parallel B_1,\dots, B_k)$, and $h_G^{k,\varepsilon}$  to graphs with vertex weights and edge costs:
$$\delta_G(A,B) = \sum_{\substack{u\in A, v\in B\\(u,v)\in E}} c_{uv}\qquad \text{and}\qquad\phi_G(P\parallel B) = \frac{\delta(P, V\setminus(P\cup B))}{w(P)}$$
Quantities $\phi_G(P_1,\dots, P_k \parallel B_1,\dots, B_k)$ and $h_G^{k,\varepsilon}$ are given by formulas (\ref{eq:buffered:expansion}) and (\ref{eq:h-G-k-eps}), respectively. We say that partition $(P_1,\dots, P_k \parallel B_1,\dots, B_k)$ is $\varepsilon$-buffered if $w(B_i) \leq \varepsilon w(P_i)$ for every $i \in[k]$.

Note that the definitions of $L_G$, $\delta_G$, $\phi_G$, and $h_{G}^{k,\varepsilon}$ are consistent with those for regular graphs with unit vertex weights and unit edge costs.
As a side note, we observe that the definition of $\Lap_G$ coincides with the definition of the normalized Laplacian in the standard Cheeger inequality for non-regular graphs with edge costs. Note that in that inequality, vertex weights are defined as $w_u = \sum_{v:(u,v)\in E} c_{uv}$. In contrast to the standard Cheeger inequality, our variant holds for  arbitrary vertex weights and edge costs.

\begin{theorem}\label{thm:main-weighted}
Let $G = (V, E, w, c)$ be a graph with positive weights $w_u > 0$ and edge costs $c_{uv} > 0$, $\varepsilon \in [0,1)$, $\delta \in (0,1)$, and $k\geq 2$ be an integer. Assume that $\max_u w_u \leq \varepsilon w(V) /(3k)$.
Then
\begin{equation}
  h_G^{k,\varepsilon} \leq \frac{\kappa(\delta) \log k}{\varepsilon} 
   \cdot \lambda_{\lfloor(1+\delta)k\rfloor}(\Lap_G),
\end{equation}
where $\kappa(\delta)$ is a function that depends only on $\delta$.
Furthermore, there is a randomized polynomial-time algorithm that given $G$ finds an $\varepsilon$-buffered $k$-partitioning $(P_1, \dots, P_k\parallel B_1, \dots, B_k)$ with $\phi_G(P_1, \dots, P_k \parallel B_1, \dots, B_k) \le \frac{\kappa(\delta) \log k}{\varepsilon}  \lambda_{\lfloor(1+\delta)k\rfloor}(\Lap_G)$. 
\end{theorem}

This new generalization with vertex weights and edge costs is crucial for the pseudoapproximation guarantees for the buffered versions of Balanced Cut (Theorem~\ref{thm:buffer-balanced}) and Balanced $k$-way partitioning (Theorem~\ref{cor:approx-k-balanced}) that were mentioned earlier.  



\subsection{Buffered Cheeger's inequality for \texorpdfstring{$k=2$}{k=2}}
For $k=2$, we provide an alternative slightly simpler variant of buffered Cheeger's inequality. We give a polynomial-time algorithm that partitions $V$ into three disjoint sets: parts $S$, $T$, and buffer $B$, satisfying $S\cup T \cup B = V$ and $|B|\le \varepsilon \min(|S|, |T|)$. The
buffered expansion of $S$ and $T$, defined as $\delta(S,T)/\min(w(S),w(T))$ is at most $O(\lambda_2/ \varepsilon)$ (see Proposition~\ref{prop:cheeger} for details). 

We provide a self-contained proof of this simpler result for $k=2$ in Section~\ref{sec:warm-up}. We remark that this result  coupled with Lemma~\ref{lem:partial-to-complete} from this paper and
Theorem 4.6 
from the paper by Lee, Oveis-Gharan, and Trevisan~\cite{LeeOvT12} 
already yields weak versions of our main results (Theorems~\ref{thm:main} and~\ref{thm:main-weighted}) where $O(\log k)$ is replaced with $O(\log^2 k)$. 
This extra logarithmic factor is a large loss in the context of graph partitioning problems, and this is analogous to the weaker higher order Cheeger inequality obtained in \cite{LeeOvT12} by combining Theorem 4.6 of \cite{LeeOvT12} with the standard Cheeger inequality for $k=2$.\footnote{The stronger bound of Theorem 4.1 in \cite{LeeOvT12} avoids Theorem 4.6.} To get a tight bound of 
$O(\dfrac{1}{\varepsilon}\log k)$, we design a new algorithm
(see the next section for why our result is tight in both $k$ and $\varepsilon$). We give an overview of new techniques in Section~\ref{sec:overview}.

\subsection{Our result generalizes higher-order Cheeger inequalities} 
Our main result (Theorem~\ref{thm:main}) can be seen as a generalization of Cheeger's inequality~(\ref{eq:basic:cheeger}) and the higher-order Cheeger inequalitiy~(\ref{eq:basic:higherorder}). To obtain these results, we 
apply Theorem~\ref{thm:main} with $\varepsilon = \sqrt{\lambda_{\lfloor(1+\delta)k\rfloor} \log k}$. We find the largest set $P_t$ among $P_1, \dots, P_k$. We may assume that $P_t$ contains at least $\Omega(\delta n)$ vertices (see Section~\ref{sec:heavy-set-Pt} for the details).
Then we include all buffers in set $P_t$; that is, we let
$P_t' = P_t \cup \bigcup_i B_i$.
We obtain a non-buffered partition of $G$. Using that $|B_i| \leq \varepsilon |P_i|$ and $\delta(P_i, B_i) \leq d |B_i|$ (since the graph is $d$-regular), we get for $i\neq t$ (here $k' = \lfloor(1+\delta)k\rfloor$), 
$$\phi_G(P_i) = \phi_G(P_i \parallel B_i) + \frac{\delta(P_i, B_i)}{d|P_i|}
\leq \frac{c(\delta)\log k}{\sqrt{\lambda_{k'}\log k}} \lambda_{k'} +
\frac{d\cdot \sqrt{\lambda_{k'}\log k} |P_i|}{d|P_i|}
=(c(\delta) + 1) \sqrt{\lambda_{k'}\log k}.
$$
We bound $\phi_G(P_t')$ (the expansion of the updated set $P'_t$)
as follows,
$$\phi(P'_t) = \frac{\sum_{i\neq t} \delta(P_i, P'_t)}{d|P'_t|}
\leq 
\frac{\sum_{i\neq t} \phi_G(P_i) \cdot |P_i|\cdot d
}{\delta n\cdot d}
\leq
\frac{(c(\delta)+1)\sqrt{\lambda_{k'}\log k}}{\Omega(\delta n)}\sum_{i\neq t}|P_i| \leq 
\frac{c(\delta)+1}{\Omega(\delta)} \sqrt{\lambda_{k'}\log k}.
$$
Hence Theorem~\ref{thm:main} provides an alternate proof of \eqref{eq:basic:higherorder}. Furthermore, this proof suggests that the  factor of $O(\dfrac{1}{\varepsilon}\log k )$ in the upper bound of Theorem~\ref{thm:main} cannot be improved. It also shows that our inverse dependence on $\varepsilon$ is tight even for $k=2$ (as otherwise we would be able to strengthen Cheeger's inequality, which is known to be tight). 


\subsection{Connection to Robust Expansion}\label{sec:intro-robust}
Theorem~\ref{thm:main} also generalizes the Cheeger-type inequality by Kwok, Lau, and Lee~\cite{KwokLL17} that gives a bound for $\lambda_2$ in terms of \textit{robust expansion}~\cite{KLM06}. 
Let $\eta \in (0,1)$. For $S \subseteq V$, define
\begin{align}
    N_{\eta}(S) &=  \min \Big\{ |T|: T \subseteq V \setminus S, ~ \delta_G(S,T) \ge (1-\eta) \delta_G(S, V \setminus S) \Big\} \\
    \phi^{V}_{\eta}(S) &= \frac{N_{\eta}(S)}{|S|}
    \quad \text{and}\quad \phi^V_\eta(G)= \min_{S: |S| \le |V|/2} \phi^V_{\eta}(S)
\end{align}
In other words, $\phi^{V}_{\eta}(S)$ is the vertex expansion of set $S$ after we remove an $\eta$ fraction of the edges leaving $S$ in the optimal way (which minimizes the vertex expansion of $S$ in the remaining graph). Quantity $\phi^{V}_{\eta}(S)$ is less sensitive to additions of a small number of edges to graph $G$ than the standard vertex expansion. For that reason,  $\phi^{V}_{\eta}(S)$ is called the \textit{robust vertex expansion} of $G$.
Kwok, Lau, and Lee~\cite{KwokLL17} proved the following result for $\eta=1/2$.
\begin{theorem}[see Theorem 1 in \cite{KwokLL17}]\label{thm:kwok} 
$
\lambda_2 = \Omega\Big( h_G \cdot \phi^V_{\sfrac{1}{2}}(G) \Big).
$
\end{theorem}
\noindent The following generalization of Theorem~\ref{thm:kwok} is an immediate corollary of Theorem~\ref{thm:main} (see Appendix~\ref{app:robustexp} for a proof).
\begin{corollary}\label{corr:robustexp}
For every $\eta \in (0,1)$ we have
$
    \lambda_2 =\Omega\Big(\eta \cdot h_G \cdot \phi^V_{\eta}(G) \Big).
$
\end{corollary}
We remark that Theorem~\ref{thm:kwok} is related to the case $k=2$ in Theorem~\ref{thm:main}. 

\subsection{Lower Bounds} 
We also prove a lower bound on $h_G^{k,\varepsilon}$, which is linear in $\lambda_k$.
\begin{theorem}\label{thm:lower-bound-hkG}
For every $d$-regular graph $G$, integer $k\geq 2$, and $\varepsilon > 0$, we have, 
$$h^{k,\varepsilon}_G \geq \frac{\lambda_k - \varepsilon}{2}.$$
\end{theorem}
We remark that the additive dependence on $\varepsilon$ in the above lower bound (Theorem~\ref{thm:lower-bound-hkG}) is unavoidable even when $k=2$.\footnote{For the tight example, consider two cliques on vertex sets $A$ and $B$ of size $(1+\varepsilon) n/2$ each, with overlap of $|A \cap B|=\varepsilon n$ vertices and with no edges between $A \setminus B$ and $B \setminus A$. Some of the edges incident on $A \cap B$ are resampled to ensure (approximate) regularity.  While $h_G^{2,\varepsilon}=0$, it is easy to show that $\lambda_2 = \Omega(\varepsilon)$.} 
This is useful to derive a lower bound on the optimal buffered expansion $h_G^{k, \varepsilon}$; moreover in conjunction with the upper bound (applied with a larger $\varepsilon'$), one can also get a bicriteria approximation for buffered $k$-way partitioning.\footnote{\label{foot:bicriteria}
%
Specifically, for any $\varepsilon \in [0,1), \delta \in (0,1)$, and $\varepsilon'>\varepsilon$, our algorithm given a graph $G$ finds an $\varepsilon'$-buffered $k$-partitioning $(P_1, \dots, P_k\parallel B_1, \dots, B_k)$ with $\phi_G(P_1, \dots, P_k \parallel B_1, \dots, B_k) \le c(\delta) \log k \cdot \big(h_G^{\lfloor k(1+\delta)\rfloor, \varepsilon}+\varepsilon\big)/\varepsilon'$, where $c(\delta)>0$ is a constant that only depends on $\delta$. 
}

\subsection{Overview and Organization}\label{sec:overview}
We start with proving a weaker version of our main result (Theorem~\ref{thm:main}) for $k=2$ in Section~\ref{sec:warm-up}. This proof is significantly simpler than the general proof but nevertheless illustrates why we get a linear dependence on $\lambda_k$ rather than a square-root dependence in our Cheeger-type inequality. In the proof, we use the thresholding idea from the proof of the standard Cheeger inequality but add an extra twist --  use two thresholds instead of one. First, we compute the eigenvector $u$ corresponding to the second smallest eigenvalue $\lambda_2$ of the normalized Laplacian $L_G$ of $G$. Let $u(i)$ be the $i$-th coordinate of $u$. 
Recall that in the proof of Cheeger's inequality, we put each vertex $i$ either in $S$ or in $T$, depending on whether $u(i)^2 \geq \tau$ or $u(i)^2 < \tau$ for an appropriately chosen threshold $\tau$. To prove our inequality for $k=2$, we use two thresholds $\tau$ and $(1+\varepsilon)\tau$ and, loosely speaking (see Section~\ref{sec:warm-up} for the precise description), put $i$ in $T$, $B$, $S$ depending on whether $u(i)^2$ lies in $(-\infty, \tau]$, $(\tau, (1+\varepsilon)\tau)$, or $[(1+\varepsilon)\tau,\infty)$, respectively.

In the subsequent sections, we prove the main result i.e., Theorem~\ref{thm:main} for arbitrary $k$. Recall the definition of the spectral embedding of graph $G$, which we use in our proof. Let $x_1, \dots, x_{k'}$ be the eigenvectors of $L_G$ corresponding to the $k' = \lfloor (1+\delta)k\rfloor$ smallest eigenvalues. Note that the coordinates of vectors $x_i$ are indexed by vertices $u$; denote the coordinate with index $u$ by $x_i(u)$. The spectral embedding maps vertex $u$ to vector $\bar u \in {\mathbb R}^{k'}$ with coordinates $x_1(u), \dots, x_{k'}(u)$. We compute the spectral embedding. And now our goal is to partition vectors $\bar u$ (so that the corresponding buffered partition satisfies the desired properties). To do so, we introduce a new technical tool -- \textit{orthogonal separators with buffers} -- for partitioning sets of vectors. 

Given a set of unit vectors, the orthogonal separator procedure generates three (disjoint) random sets -- set $X$ (called an orthogonal separator) and its two buffers $Y$ and $Z$ -- such that 
\begin{enumerate}
\item if $u\in X$ and $v$ is close to $u$ then $v$ is in $X\cup Y\cup Z$ with high probability 
\item if vectors $u$ and $v$ are far apart, then it is unlikely that both of them are in $X$
\item $|Y|,|Z|$ are at most $\varepsilon |X|$ in expectation
\end{enumerate}

(See Theorems~\ref{thm:orth-sep} and~\ref{thm:cor:orth-sep-main} for details.) Orthogonal separators with buffers provide a basic building block for constructing buffered partitionings. We repeatedly apply the orthogonal separator procedure to normalized vectors $\psi(\bar u) = \frac{\bar u}{\|\bar u\|}$ and obtain subsets $X_t$ and their buffers $Y_t,Z_t$. Merging the obtained sets and filtering/thresholding them based on the lengths of vectors $\bar u$, we obtain a \textit{partial} buffered partitioning. This partitioning has all the desired properties except that it does not necessarily cover the entire vertex set $V$. While we do not provide any details on how this step works in this overview, note that we use item 1 to argue that the buffered expansion of each set $P_i$ is small, item 2 to argue that the obtained sets are not too large and thus there are at least $k$ sets in the partitioning, and item 3 to argue that $|B_i| \leq \varepsilon |P_i|$.

Note that orthogonal separators with buffers generalize (non-buffered) orthogonal separators introduced by Chlamtac, Makarychev, and Makarychev~\cite{CMM2} and used in a number of SDP-based approximation algorithms for graph partitioning problems. An analog of Theorem~\ref{thm:cor:orth-sep-main} for (non-buffered) orthogonal separators was first proved by
Bansal, Feige, Krauthgamer, Makarychev, Nagarajan, Naor, and Schwartz~\cite{bansal2014min} (see also~\cite{LM14}). Our high level  approach follows the paper by Louis and Makarychev~\cite{LM14}. However, our algorithm and its analysis substantially differ from theirs because we need to use \emph{orthogonal separators with buffers} and keep track of the buffers between clusters. Also, our algorithm uses a spectral embedding while the algorithm by Louis and Makarychev~\cite{LM14} uses an embedding obtained from an SDP relaxation, which 
imposes additional constraints on vectors.

We prove some useful claims about the spectral embedding in Section~\ref{sec:spectral}.
We define orthogonal separators with buffers and present the main theorem about them (Theorem~\ref{thm:cor:orth-sep-main}) in Section~\ref{sec:orth-sep}. We prove Theorem~\ref{thm:cor:orth-sep-main} in Section~\ref{sec:os-proof}. We show how to obtain a partial buffered clustering in Section~\ref{sec:partial-partition}. Finally, in Section~\ref{sec:partial-tocomplete}, we show how to obtain a true buffered partitioning. 

The proof of the Cheeger inequality for graphs with arbitrary vertex weights and edge costs (Theorem~\ref{thm:main-weighted}) is almost identical to that of Theorem~\ref{thm:main}. In order to simplify the exposition, we only present the proof of Theorem~\ref{thm:main}. The same proof with minimal changes works in the general case. Instead of presenting essentially the same proof again, we
give a black box reduction from Theorem~\ref{thm:main} to Theorem~\ref{thm:main-weighted} in Appendix~\ref{sec:reduction}. The reduction however may significantly increase the running time of the algorithm. We stress that the algorithm from Theorem~\ref{thm:main} also works with weighted graphs.

The other sections and appendices are organized as follows. In Section~\ref{app:robustexp}, we show that Theorem~\ref{thm:main} implies Corollary~\ref{corr:robustexp}, which we discussed in Section~\ref{sec:intro-robust}.
In Section~\ref{sec:heavy-set-Pt}, we prove a technical claim about $\varepsilon$-buffered partitions.
In Section~\ref{sec:lb-and-approx}, we prove a lower bound on $h_G^k$ for unbuffered partitions of graphs $G$ with vertex weights and edge costs. Combining this lower bound with Theorem~\ref{thm:main-weighted}, we get a pseudo-approximation algorithm for the Sparsest $k$-way Partitioning problem (Theorem~\ref{thm:approx-sparsest-k-partitioning}).
In Section~\ref{sec:buf-balanced-cut}, we present our pseudo-approximation algorithm for the Buffered Balanced Cut problem.
In Section~\ref{sec:lowerbound}, we prove Theorem~\ref{thm:lower-bound-hkG} (a lower bound on $h_G^{k,\varepsilon}$ discussed above). In Section~\ref{sec:gaussian}, we give a few useful estimates on the Gaussian distribution, which we use throughout the paper.

\paragraph{Other related work.} Clustering with vertex deletion and duplication has been studied in other context as well. We refer the reader to the following recent results:
Filtser and Le~\cite{filtser2021clan},
Haeupler, Hershkowitz, and Zuzic~\cite{haeupler2021tree},
Filtser~\cite{filtser2022hop}.
\section{Warm up: Cheeger's Inequality with a Buffer for \texorpdfstring{$k=2$}{k=2}}\label{sec:warm-up}

As a warmup, we provide a self-contained proof of a weaker version of  Theorem~\ref{thm:main} for $k=2$. Here, we will consider cuts $(S,T)$ with a common buffer $B$ (instead of disjoint buffers for $S$ and $T$). Such cuts consist of three disjoint sets $S$, $T$, and $B$ that partition the set of vertices $V$ into three groups. We will refer to such a partition as $(S,T\parallel B)$. While there are many new ideas needed to obtain Theorem~\ref{thm:main} in full generality, this simpler setting already demonstrates how one can leverage buffers to obtain an improved upper bound. 


\begin{proposition}\label{prop:cheeger}
Let $\varepsilon \in (0,1/4)$.
Consider any graph $G=(V,E)$ with positive vertex weights $w_u > 0$ and edge costs $c_{uv} > 0$.  Let $\lambda_G$ be the second smallest eigenvalue of $L_G = D_w^{-1/2}\tilde L_G D_w^{-1/2}$, the normalized Laplacian of $G$. Then, in polynomial time we can find three disjoint sets $S, B, T$ with $S\cup B\cup T = V$, $w(S) \le w(T)$ and $w(B) \le \varepsilon w(S)$ such that  
$$
\phi_G(S,T\parallel B) = \frac{|\delta(S,T)|}{w(S)}
\leq 4\Big(1+\frac{2}{\varepsilon}\Big)\lambda_G.
$$
\end{proposition}


\begin{proof} The proof follows the same general strategy as the standard proof of the Cheeger inequality. 
We show how to find a distribution over (buffered) partitions $(S, B, T)$ in the graph $G$, by thresholding the second eigenvector of $L_G$, such that:
$$
\text{(I) } \E|\delta(S,T)| \leq (1+1/\varepsilon)\lambda_G\cdot  \E[w(S)] \quad\text{ and }\quad
\text{(II) } \E[w(B)] \le \varepsilon \E[w(S)].$$ 

\noindent 
The first condition gives an upper bound on the expected number of (non-buffered) edges crossing the cut, while the second condition gives a bound on the expected size of the buffer.  A simple probabilistic argument (see Lemma~\ref{lem:both-hold}) allows us to conclude that there exists a single buffered threshold cut that simultaneously satisfies both the properties (with some slack).

Consider the spectrum of 
matrix $L_G= D_w^{-1/2}\tilde L_G D_w^{-1/2}$. The first eigenvector of the non-normalized Laplacian $\tilde L_G$ is the vector of 
all ones denoted by $\ONE$. Its eigenvalue is 0. In other words, $\tilde L_G\ONE = 0$.
Consequently, 
$L_G
(D_w^{1/2}\ONE) = D_w^{-1/2}\tilde L_G \ONE = 0$. Hence, $D_w^{1/2}\ONE$ is the first eigenvector of 
$L_G$. Let $y$ be an eigenvector of $L_G$  corresponding to the second eigenvalue $\lambda_G=\lambda_2$ of $L_G$. Then,
$y \perp D_w^{1/2}\ONE$ and
\begin{equation}\label{eq:y-LG-y-bound}
\langle y, L_G y\rangle  = \langle y, D_w^{-1/2}\tilde L_G D_w^{-1/2} y\rangle = \lambda_G \|y\|^2.    
\end{equation} 
Let $v = D_w^{-1/2}y$. Then, we have $v \perp D_w\ONE$ 
(because 
$\langle v, D_w\ONE\rangle = 
\langle y, D_w^{1/2}\ONE\rangle = 0$)
and
\begin{equation}\label{eq:v-LG-v-bound}
\langle v, \tilde L_G v\rangle = 
\langle D_w^{-1/2} y, \tilde L_G D_w^{-1/2} y\rangle = \lambda_G \|y\|^2 =
\lambda_G \|D_w^{1/2}v\|^2.
\end{equation}  

\paragraph{Step 1. Splitting the vector.} 
For technical reasons, we need to split vector $v$ into two vectors $v_+$ and $v_-$ such that the vertex weight of non-zero coordinates in each vector is at most $w(V)/2$,
$$
w(\{i: v_+(i) > 0\}) \leq w(V)/2; \quad w(\{i: v_-(i) > 0\}) \leq w(V)/2.
$$
We do this by following a standard trick that is often used in the proof of Cheeger's inequality. 
Let $z$ denote the smallest coordinate value in the vector $v$ such that the total vertex weight of coordinates with a value greater than $z$ in vector $v$ is at most $w(V)/2$, i.e. 
$$
w(\{i: v(i) >z\}) \leq w(V)/2; \quad w(\{i: v(i) < z\}) \leq w(V)/2.
$$ 

Then 
we shift the entire vector $v$ by $z$ and get 
$v' = v - z\ONE$. 
Since $\tilde L_G\ONE = 0$ and $v \perp D_w\ONE$, we have 
$$
\langle v', \tilde L_G v' \rangle
= 
\langle v, \tilde L_G v \rangle
-
\underbrace{
2z\langle v, \tilde L_G 
\ONE \rangle}_{=0}
+ 
\underbrace{z^2 \langle \ONE, \tilde L_G \ONE \rangle}_{=0} 
\eqbecause{by (\ref{eq:v-LG-v-bound})} \lambda_G
\|D_w^{1/2}v\|^2\leq \lambda_G \|D_w^{1/2}v'\|^2.
$$
The last inequality holds because 
$$
\|D_w^{1/2}v'\|^2 = 
\|D_w^{1/2}v\|^2 
+
z^2 \|D_w^{1/2}\ONE\|^2 
-2z\langle 
D_w^{1/2}v,
D_w^{1/2}\ONE
\rangle = 
\|D_w^{1/2}v\|^2 
+
z^2 \underbrace{\|D_w^{1/2}\ONE\|^2}_{\geq 0} 
-2z
\underbrace{\langle 
v,D_w\ONE
\rangle}_{=0}.
$$
We now split the vector $v'$ into two vectors $v_+, v_-$ with disjoint supports as follows:
$$v_+ (i)= \begin{cases}
v(i) - z,& \text{if } v(i) \geq z;\\
0,& \text{otherwise},
\end{cases} \qquad
 v_- (i)= \begin{cases}
0,& \text{if } v(i) \geq z;\\
v(i)- z,& \text{otherwise}.
\end{cases}
$$
\begin{claim}\label{cl:v-plus-v-minus-new}
For $u = v_+$ or $u= v_-$, we have  $u \neq 0$ and $\langle u, \tilde L_G u \rangle
\leq \lambda_G \|D_w^{1/2}u\|^2$.
\end{claim}
\begin{proof}
Vectors $D_w^{1/2}v_+$ and $D_w^{1/2}v_-$ are orthogonal because their supports are disjoint (note: $D_w^{1/2}$ is a diagonal matrix). All coordinates of $D_w^{1/2} v_+$ are non-negative, and all coordinates of
$D_w^{1/2}v_-$ are non-positive. Thus,
$
\|D_w^{1/2} v_{+}\|^2 + \|D_w^{1/2} v_{-}\|^2=  \|D_w^{1/2}(v_{+} + v_{-})\|^2
=\|D_w^{1/2}v'\|^2$ and 
$$
\langle v', \tilde L_G v'\rangle = 
\langle v_+, \tilde L_G v_+\rangle +  \langle v_-, \tilde L_G v_-\rangle +  \underbrace{2\langle v_-, \tilde L_G v_+\rangle}_{\geq 0}
\geq \langle v_+, \tilde L_G v_+\rangle +  \langle v_-, \tilde L_G v_-\rangle.$$
The last inequality holds because all off diagonal entries in $\tilde L_G$ are non-positive;  $v_+(i)v_-(j)\leq 0$ for all $i\neq j$; and $v_+(i)v_-(i) = 0$. We have
$$\langle v_+, \tilde L_G v_+\rangle +  \langle v_-, \tilde L_G v_-\rangle \leq \langle v', \tilde L_G v'\rangle \leq
\lambda_G \|D_w^{1/2} v'\|^2 = \lambda_G (\|D_w^{1/2} v_{+}\|^2 + \|D_w^{1/2} v_{-}\|^2).$$
Thus, for $u = v_+$ or $u=v_-$ the desired inequality holds.
\end{proof}

Let $u$ be as above. We assume without loss of generality that $\|u\|_\infty = \max_u |u(i)|=1$ (if $\|u\|_\infty \neq 1$, we divide $u$ by  $\|u\|_\infty$). 
Next, we show that there exists an $\varepsilon$-buffered partition with small expansion by thresholding on this vector $u$. 


\paragraph{Step 2. Random Thresholding with Buffers.}

Pick a random threshold $t \in [0,1]$  uniformly distributed in $[0,1]$ and define sets $S$, $T$, and buffer $B$ as follows:
\begin{align}\label{eq:cheeger2:defs}
  S&=\{i: u(i)^2 > t\}\\ T&=\big\{i: u(i)^2 \leq \sfrac{t}{(1+\varepsilon)}\big\}, \label{def:T-k-2} \\
   B&= V \setminus (S\cup T) = 
\big\{i: \sfrac{t}{(1+\varepsilon)} < u(i)^2 \le t \big\}. 
\end{align}
Note that $B \cup S= \{i: u(i)^2  > \sfrac{t}{(1+\varepsilon)}\}$. Since $t$ is picked uniformly from $[0,1]$ and $\|u\|_\infty =1$, we have 
$$\E[w(S)]= \sum_{i=1}^n w_i \Pr\{i\in S\} = \sum_{i=1}^n w_i \cdot u(i)^2 = \|D_w^{1/2}u\|^2,$$
and
\begin{align}
    \E[w(B\cup S)] &= \sum_{i=1}^n w_i\cdot \min((1+\varepsilon)|u(i)|^2,1) \leq (1+\varepsilon)\|D_w^{1/2}u\|^2. 
\end{align}
Thus,
$\E[w(B)] \leq \varepsilon \|D_w^{1/2}u\|^2 = \varepsilon \E[w(S)]$, as stated in  Equation~(II).

By our choice of $z$, the weight of vertices with positive values in $u$ is at most $w(V)/2$. Since $S$ contains a subset of vertices with positive values in $u$, we have $w(S) \leq w(V)/2$.


Note that for every edge $(i,j)$ from $S$ to $T$, we have $u(i)^2 > t > t/(1+\varepsilon) \geq u(j)^2$. Thus, for all edges $(i,j) \in \delta(S,T)$, we have: (a) $i\in S, j\in T$ if $u(i)^2 > u(j)^2$ and (b) $i\in T, j\in S$ if $u(i)^2 < u(j)^2$. 
Now consider an edge $(i,j) \in E$ with $u(i)^2 > u(j)^2$.
The probability that $(i,j)\in \delta(S,T)$  equals
\begin{align*}
\Pr\{(i,j)\in \delta(S,T)\}&=
\Pr\{i\in S; j\in T\} = 
\Pr\{t \leq u(i)^2\;\;\&\;\; t \geq (1 + \varepsilon) u(j)^2\}\\
&= \max\{u(i)^2 - (1 + \varepsilon) u(j)^2, 0\}.
\end{align*}
To bound the right side, we use the following simple claim.
\begin{claim}\label{cl:a-minus-b}
For all $\varepsilon>0$ and all real numbers $a$ and $b$, we have
$$a^2 - (1+\varepsilon) b^2 \leq (1+\nicefrac{1}{\varepsilon}) (a-b)^2.$$
\end{claim}
\begin{proof}
 If $b=0$, then the inequality holds. Assume, that $b\neq 0$. Divide both sides by $b^2$ and denote $\lambda = a/b$. We need to show that
$(1+\nicefrac{1}{\varepsilon})(\lambda-1)^2
 - \big(\lambda^2 -(1+\varepsilon)\big) \geq 0$.
 Write,
 \begin{align*}
 (1+\nicefrac{1}{\varepsilon})(\lambda-1)^2
 - \big(\lambda^2 -(1+\varepsilon)\big) &= 
 \nicefrac{1}{\varepsilon} \lambda^2 - 2(1+\nicefrac{1}{\varepsilon})\lambda + (\sqrt{\varepsilon} + \nicefrac{1}{\sqrt{\varepsilon}})^2\\
 &= (\nicefrac{\lambda}{\sqrt{\varepsilon}} - (\sqrt{\varepsilon} + \nicefrac{1}{\sqrt{\varepsilon}}))^2 \geq 0.\qedhere
 \end{align*}
\end{proof}

\noindent Hence from the above Claim~\ref{cl:a-minus-b}, we have
$$
\Pr\{i\in S; j\in T\} \leq (1+\nicefrac{1}{\varepsilon})(u(i) - u(j))^2.$$
By linearity of expectation, 
\begin{multline*}
\E|\delta(S,T)| \leq (1+\nicefrac{1}{\varepsilon})
\sum_{\substack{(i,j)\in E\\u(i)^2 > u(j)^2}} c_{ij}(u(i) - u(j))^2 \stackrel{\text{\tiny by (\ref{eq:non-normalized-L})}}{=} 
 (1+\nicefrac{1}{\varepsilon}) \langle u, \tilde L_G u\rangle \leq \\
\leq  (1+\nicefrac{1}{\varepsilon})\lambda_G \|D_w^{1/2}u\|^2\leq (1+\nicefrac{1}{\varepsilon})\lambda_G \cdot \E[w(S)].
\end{multline*}
We bounded $\langle u, \tilde L_G u\rangle$ using Claim~\ref{cl:v-plus-v-minus-new} (cf. Equation~(\ref{eq:v-LG-v-bound})).
Thus, this distribution over buffered partitions $(S,T \parallel B)$ 
satisfies Equation (I).
Since (I) and (II) both hold, we can use Lemma~\ref{lem:both-hold} (see below) to conclude that there exists a cut $(\hat S,\hat T)$ with buffer $\hat B$ for which 
$$ 
|\delta(\hat S,\hat T)| \leq 2(1+\nicefrac{1}{\varepsilon})\lambda_G \cdot w(\hat S),\text{ and }
w(\hat B) \leq  2\varepsilon \cdot w(\hat{S}).
$$
For this cut $(\hat S,\hat T)$ with buffer $\hat B$, we have
$$ 
\frac{|\delta(\hat S,\hat T)|}{w(\hat S)} \leq \frac{2(1+\nicefrac{1}{\varepsilon})\lambda_G \cdot w(\hat S)}{w(\hat S)} = 2(1+\nicefrac{1}{\varepsilon})\lambda_G.
$$
By (\ref{eq:cheeger2:defs}) and (\ref{def:T-k-2}), we have $\hat{S} \subseteq \{i: u(i)^2 > 0\}$ and $\hat{T} \supseteq \{i: u(i)^2 = 0\}$. Thus
$w(\hat{T}) \leq w(\{i: u(i)^2 > 0\}) \leq w(V)/2$ and 
$w(\hat{T}) \geq w(\{i: u(i)^2 = 0\}) = w(V) -  w(\{i: u(i)^2 > 0\}) \geq w(V)/2$. Therefore,
$w(\hat{T}) \leq w(\hat{S})$. We conclude that 
$$\frac{|\delta(\hat S,\hat T)|}{w(\hat T)}\leq \frac{|\delta(\hat S,\hat T)|}{w(\hat S)} \leq2(1+\nicefrac{1}{\varepsilon})\lambda_G.$$
We obtain the desired result for $\varepsilon' = 2\varepsilon$.
To finish the proof, it remains to show Lemma~\ref{lem:both-hold}. 

\end{proof}


\begin{lemma}\label{lem:both-hold}
    For any $m \geq 2$, consider $m$ arbitrary jointly distributed non-negative random variables $X_1,\dots,X_{m-1}$ and $Y$. Suppose that for every $i = 1,\dots, m-1$, $\E[X_i] \leq \alpha_i \E[Z]$. Then, 
    \begin{equation}\label{eq:pr-both-hold}
        \Pr\{X_i \leq 2\alpha_i Y, \quad \forall i \in [m-1]\} > 0.
    \end{equation}
\end{lemma}

\begin{proof}
    Consider a new random variable $Z = \sum_{i=1}^{m-1} \frac{X_i}{(m-1)\alpha_i}$. By the linearity of expectation, we have
    $$
    \E[Y] \geq \frac{1}{m-1}\sum_{i=1}^{m-1} \frac{\E[X_i]}{\alpha_i} = \E[Z].
    $$
    This implies that $\Pr\{Y\geq Z\} > 0$; otherwise we would have $\E[Y] < \E[Z]$. If $Y \geq Z$, then we have for every $i = 1, \dots, m-1$, $X_i \leq (m-1)\alpha_i Y$. Therefore,
inequality (\ref{eq:pr-both-hold}) holds.
\end{proof}

\section{Orthogonal Separators with Buffers}\label{sec:orth-sep}

In this section, we introduce orthogonal separators with buffers. We will prove Theorems~\ref{thm:orth-sep}, \ref{thm:cor:orth-sep-main}, and~\ref{thm:orth-sep-two-buffers} in Section~\ref{sec:os-proof}. In these theorems, we provide randomized procedures to generate orthogonal separators with buffers in a set of unit vectors $U$ in $\bbR^d$. 
In the next section, we will use the procedure in Theorem~\ref{thm:orth-sep-two-buffers} to create a partial partitioning. We first use spectral embedding to map each vertex $u\in V$ to a vector $\bar{u} \in \bbR^k$. We will run this procedure on normalized vectors $\psi(\bar{u}) = \bar{u}/ \|\bar{u}\|$ for all vertices $u \in V$. We first give the definition of the orthogonal separator with one buffer. 

\begin{definition}
    Consider a finite set $U$ of unit vectors in $\bbR^d$. A distribution over two disjoint subsets of $U$ is an $m$-orthogonal separator with an $\varepsilon$-buffer, distortion $\calD$, separation radius $R$, and probability scale $\alpha$ if the following conditions hold for two subsets $X,Y \subseteq U$ chosen according to this distribution:
    \begin{enumerate}
        \item For all $\bar u \in U$, $\Pr\{\bar u \in X\} = \alpha$.
        \item For all $\bar u \in U$, $\Pr\{\bar u \in Y\} \leq \varepsilon\alpha$.
        \item For all $\bar u, \bar v \in U$ with $\|\bar u - \bar v\| \geq R$, $\Pr\{\bar v \in X \given \bar u \in X\} \leq \frac{1}{m}$.
        \item For all $\bar u, \bar v \in U$, $\Pr\{\bar v\notin X\cup Y \mid \bar u \in X\} \leq \calD\, \|\bar u - \bar v\|^2$.
    \end{enumerate}
    We  call $X$ an orthogonal separator and $Y$ its buffer.
\end{definition}

In this definition, conditions 1 and 2 restrict the size of an orthogonal separator and its buffer respectively. Condition 3 requires that for every pair of vectors $\bar u,\bar v \in U$, if $\bar u, \bar v$ are almost orthogonal, then vectors $\bar u,\bar v$ are separated by $X$ with high probability. Condition 4 upper bounds the probability that vectors $\bar u, \bar v$ are separated by the orthogonal separator $X$ with a buffer $Y$. In the following theorem, we show there exists such an orthogonal separator with one buffer. The construction of the orthogonal separator with one buffer and its proof is in Section~\ref{sec:os-proof}.

\begin{theorem}\label{thm:orth-sep}
There exists a randomized polynomial-time procedure that given a finite set $U$ of unit vectors in $\bbR^d$ and positive parameters $\varepsilon \in (0,1), m \geq 3, R \in (0,2)$, returns an $m$-orthogonal separator with an $\varepsilon$-buffer with distortion $\calD = O_R(\nicefrac{1}{\varepsilon}\;\log m)$, separation radius $R$, and probability scale $\alpha \geq O_R(1/poly(m))$.
\end{theorem}


In the above theorem, we show that if vectors $\bar u$ and $\bar v$ are far apart, then they are both contained in $X$ with a small probability. Suppose that every point $\bar u$ has a certain weight or measure $\mu(\bar u)$. We now show that by slightly altering the distribution of $X$ and $Y$,
we can guarantee that the measure of every $X$ is not much larger than the measure of the heaviest ball of radius $R$ (see item 3 below for details).

\begin{definition}
    Consider a finite set $U$ of unit vectors in $\bbR^d$ equipped with a measure $\mu$. A distribution over two disjoint subsets of $U$ is an $\delta$-orthogonal separator with an $\varepsilon$-buffer, distortion $\calD$, separation radius $R$, and probability scale $\alpha$ if the following conditions hold for two subsets $X,Y \subseteq U$ chosen according to this distribution:
    \begin{enumerate}
        \item For all $\bar u \in U$, $\Pr\{\bar u \in X\} = \alpha$.
        \item For all $\bar u \in U$, $\Pr\{\bar u \in Y\} \leq \varepsilon\alpha$.
        \item $\min_{\bar u \in X} \mu(X\setminus \Ball(\bar u, R))\leq \delta \mu(U)$ (always).
        \item For all $\bar u, \bar v \in U$, $\Pr\{\bar v\notin X\cup Y \mid \bar u \in X\} \leq \calD\, \|\bar u - \bar v\|^2$.
    \end{enumerate}
\end{definition}

\begin{theorem}\label{thm:cor:orth-sep-main}
There exists a randomized procedure that given a finite set $U$ of unit vectors in $\bbR^d$ equipped with a measure $\mu$ and positive parameters $\varepsilon \in (0,1), \delta \leq 2/3, R \in (0,2)$, returns an $\delta$-orthogonal separator with an $\varepsilon$-buffer with distortion $\calD = O_R(\nicefrac{1}{\varepsilon}\;\log \nicefrac{1}{\delta})$, separation radius $R$, and probability scale $\alpha \geq O_R(1/poly(m))$.
\end{theorem}


By using the orthogonal separator with one buffer above, we can find a buffered partitioning of the graph with buffered expansion in Theorem~\ref{thm:main}, but buffers $B_i$ may overlap. To get disjoint buffers as in Theorem~\ref{thm:main}, we use the orthogonal separator with two buffers defined as follows.  

\begin{definition}
    Consider a finite set $U$ of unit vectors in $\bbR^d$ equipped with a measure $\mu$. A distribution over three disjoint subsets of $U$ is an $\delta$-orthogonal separator with two $\varepsilon$-buffers, distortion $\calD$, separation radius $R$, and probability scale $\alpha$ if the following conditions hold for three disjoint subsets $X,Y,Z \subseteq U$ chosen according to this distribution:
    \begin{enumerate}
        \item For all $\bar u \in U$, $\Pr\{\bar u \in X\} = \alpha$.
        \item For all $\bar u \in U$, $\Pr\{\bar u \in Y\} \leq \varepsilon\alpha$ and $\Pr\{\bar u \in Z\} \leq \varepsilon \alpha$.
        \item $\min_{\bar u \in X} \mu(X\setminus \Ball(\bar u, R))\leq \delta \mu(U)$ (always).
        \item For all $\bar u, \bar v \in U$, $\Pr\{\bar v\notin X\cup Y \mid \bar u \in X\} \leq \calD\, \|\bar u - \bar v\|^2$, and \\ $\Pr\{\bar v\notin X\cup Y\cup Z \mid \bar u \in X \cup Y\} \leq \calD\, \|\bar u - \bar v\|^2$.
    \end{enumerate}
\end{definition}

In the following theorem, we slightly modify the procedure above to get orthogonal separators with two buffers.

\begin{theorem}\label{thm:orth-sep-two-buffers}
    There exists a randomized procedure that given a finite set $U$ of unit vectors in $\bbR^d$ equipped with a measure $\mu$ and positive parameters $\varepsilon \in (0,1), \delta \leq 2/3, R \in (0,2)$, returns an $\delta$-orthogonal separator with two $\varepsilon$-buffers with distortion $\calD = O_R(\nicefrac{1}{\varepsilon}\;\log \nicefrac{1}{\delta})$, separation radius $R$, and probability scale $\alpha \geq O_R(1/poly(m))$.
\end{theorem}


\section{Partial Partitioning}\label{sec:partial-partition}
In this section, we give an algorithm for finding a partial $\varepsilon$-buffered partitioning $(P_1,B_1),\dots, (P_{k'}, B_{k'})$ of $G$. This partitioning satisfies all the properties of the partitioning from Theorem~\ref{thm:main} except the union of sets $P_i$ does not necessarily cover the entire vertex set of $G$. 
For notational convenience, we will use $k$ to denote the index of the eigenvalue that we compare the cost to. Eventually this theorem will be applied with $k=(1+O(\delta))\hat{k}$, where $\hat{k}$ is the desired number of clusters (which we denoted by $k$ in Theorem~\ref{thm:main}). 
We obtain this partial partitioning using Algorithm 1 which consists of Steps 1, 2, 3, and 4 provided in Figures~\ref{fig:step-1},
\ref{fig:step-2},
\ref{fig:step-3}, and
\ref{fig:step-4}. 

Algorithm 1 generates this partial partitioning $(P_1,B_1),\dots, (P_{k'}, B_{k'})$ with $k' \geq (1-2\delta)k$ and partitions the uncovered vertices $V \setminus \bigcup_{i \in [k']} P_i\cup B_i$ into disjoint subsets $A_i',A_i''$ for $i \in [k']$ and $R_B',R_P'$. 
We prove that these subsets $P_i,B_i,A_i',A_i''$ for $i \in [k']$ and $R_B',R_P'$ satisfy six properties given in Theorem~\ref{thm:main-partial-part} (see below).
The first three properties show subsets $P_i,B_i$ forms a partial $\varepsilon$-buffered partitioning. 
In Section~\ref{sec:partial-tocomplete}, we show how to transform this partial partitioning with 
$k'$ clusters into a true buffered partitioning of $G$ with $\hat{k}$ clusters. We combine those additional sets $A_i',A_i'', R_P',R_B'$ to get a true buffered partitioning. The properties 4, 5, and 6 in Theorem~\ref{thm:main-partial-part}  are used in Section~\ref{sec:partial-tocomplete}.

\begin{figure}
\begin{tcolorbox}
Find a spectral embedding for graph $G$:
\begin{itemize}
    \item Let $L_G$ be the normalized Laplacian matrix for $G$. 
    \item Find the top $k$ eigenvalues of $L_G$ and corresponding orthogonal unit eigenvectors $x_1,\dots,x_k \in \bbR^V$. Denote coordinate $u\in V$ of $x_i$ by $x_i(u)$.
    \item Embed each vertex $u \in V$ into $k$-dimensional vector $\bar u$ defined as follows: the $i$-th coodinate of $\bar u$ is $x_i(u)$.
\end{itemize}    
\end{tcolorbox}
    \caption{Step 1 of Partial Partitioning. At this step, the algorithm maps vertices of $G$ into vectors using the standard spectral embedding.}
    \label{fig:step-1}
\end{figure}

\begin{figure}
\begin{tcolorbox}
Let $R=\sqrt{\nicefrac{\delta}{6}}$, 
$\delta' = \nicefrac{\delta}{2k}$, and $T= \nicefrac{2}{\alpha} \ln \nicefrac{1}{\delta}$.

Set $\Sigma_0 = \varnothing$ and $\Gamma_0 = \varnothing$.

For $t=1,\dots, T$: 
\begin{itemize}
    \item Sample an orthogonal separator $X_t$ with buffers $Y_t, Z_t$ using Theorem~\ref{thm:orth-sep-two-buffers} with parameters $\varepsilon$, $R$, and $\delta'$. For convenience, we assume that $X_t$, $Y_t$, and $Z_t$ contain not vectors but the corresponding vertices of $G$.
    
    \item Let $\widetilde P_t = X_t\setminus (\bigcup_{i <t} X_i \cup Y_i \cup Z_i)$ and $\Sigma_t = \Sigma_{t-1} \cup \widetilde P_t$.
    \item Let $\widetilde B_t =
    (X_t \cup Y_t) \setminus  (\Sigma_t \cup \Gamma_{t-1})$ and $\Gamma_t = \Gamma_{t-1} \cup \widetilde B_t$.
    \item Let $R_P = V \setminus (\bigcup_{t=1}^T X_t \cup Y_t \cup Z_t)$ and $R_B = V \setminus (\Sigma_T \cup \Gamma_T \cup R_P)$.
\end{itemize}    
\end{tcolorbox}
\caption{Step 2 of Partial Partitioning. At this step, the algorithm finds a \emph{crude} partial partitioning $\{(\widetilde P_t, \widetilde B_t)\}_t$ of $V$.}
\label{fig:step-2}
\end{figure}

\begin{figure}
\begin{tcolorbox}
Let $R_P' = R_P$ and $R_B' = R_B$.

For $t=1,\cdots, T$: 
\begin{itemize}
    \item Find $r_t$ that minimizes
    $\phi_G(P_t\parallel B_t)$ subject to the constraints $|B_t|\leq C'_{\ref{thm:main-partial-part}}(\delta) \varepsilon|P_t|$, $|A_t''| \leq 10\varepsilon |P_t|$, $\delta(A_t', P_t\cup B_t) \leq \nicefrac{C''_{\ref{thm:main-partial-part}}(\delta)}{\varepsilon}\cdot\lambda_k \log k\cdot d |P_t|$
    , and $\delta_G(P_t\cup B_t, (\Sigma_T\cup R_P) \setminus \widetilde P_t) \leq \nicefrac{C''_{\ref{thm:main-partial-part}}(\delta)}{\varepsilon}\;\lambda_k \log k\cdot d|P_t|$ 
where
    \begin{align*}
P_t &= \{u\in \widetilde P_t : \|\bar u\|^2\geq r_t\}\\
B_t &= \{u\in \widetilde B_t : \|\bar u\|^2\geq r_t/(1+\varepsilon)\}\cup\{u\in \widetilde P_t : \|\bar u\|^2\in [r_t/(1 +\varepsilon),r_t]\}\\
A_t' &= \{u\in \widetilde P_t : \|\bar u\|^2 \leq  r_t/(1+\varepsilon)^2\}\\
A_t'' &= \{u\in \widetilde P_t : \|\bar u\|^2  \in (r_t/(1+\varepsilon)^2, r_t/(1+\varepsilon))\}
\end{align*}
Note that it suffices to consider $r$ in $\{\|\bar u\|^2: u\in \widetilde P_t \cup \widetilde B_t\}$. If no such $r_t$  exists, we let $P_t=\varnothing$,  $B_t=\varnothing$, $A_t' = \varnothing$, and $A_t'' = \varnothing$.

\item 
If no such $r_t$  exists, then add $\widetilde P_t$ to $R_P'$ and add $\widetilde B_t$ to $R_B'$. Otherwise, add $\widetilde B_T \setminus B_t$ to $R_B'$.  
\end{itemize}    
\end{tcolorbox}
\caption{Step 3 of Partial Partitioning. At this step, the algorithm refines the \emph{crude} partial partitioning $\{(\widetilde P_t, \widetilde B_t)\}_t$ of $V$
and obtains sets  $\{(P_t, B_t,A_t',A_t'')\}_t$.}
\label{fig:step-3}
\end{figure}

\begin{figure}
\begin{tcolorbox}
For $t=1,\cdots, T$: 

\begin{itemize}
\item
Discard all sets $P_t,B_t,A_t',A_t''$ if $P_t=\varnothing$,
or
$$\phi_G(P_t\parallel B_t) > \dfrac{C''_{\ref{thm:main-partial-part}}(\delta)}{\varepsilon}\;\lambda_k \log k,$$
where $C''_{\ref{thm:main-partial-part}}(\delta)$ is some function that depends only on $\delta$ (see Theorem~\ref{thm:main-partial-part}).
\item 
If sets $P_t,B_t,A_t',A_t''$ are discarded, then add $\widetilde P_t$ to $R_P'$ and add $\widetilde B_t$ to $R_B'$.
\end{itemize}
\end{tcolorbox}
\caption{Step 4 of Partial Partitioning. At this step, the algorithm discards all sets $(P_t,B_t)$ that do not satisfy the conditions of Theorem~\ref{thm:main-partial-part}.}
\label{fig:step-4}
\end{figure}

\newpage

\begin{theorem}\label{thm:main-partial-part}
Algorithm 1 is a polynomial-time randomized  algorithm that given a $d$-regular graph $G=(V,E)$, natural $k>1$, and positive parameters $\varepsilon,\delta\in(0,1/80)$, finds subsets $R_P',R_B'$ and
$P_i,B_i,A_i',A_i''$ of $V$ for $i\in [k']$ with $k'\geq (1-2\delta)k$ such that

\begin{enumerate}
\item 
All sets $P_i,B_i,A_i',A_i''$ and $R_P',R_B'$ are disjoint and all sets $P_i$ are nonempty, and
$$
R_P' \cup R_B' \cup \bigcup_{i=1}^{k'} P_i\cup B_i\cup A_i' \cup A_i'' = V; 
$$
\item $|B_{i}|\leq C'_{\ref{thm:main-partial-part}}(\delta)\;\varepsilon |P_i|$ for all $i\in\{1,\dots,k'\}$; and
\item $\phi_G(P_i\parallel B_i)\leq \dfrac{C''_{\ref{thm:main-partial-part}}(\delta)}{\varepsilon}\;\lambda_k \log k$, for all $i\in [k']$, 
\item
$|A_i''|\leq 10\varepsilon |P_i|$, for all $i\in [k']$;
\item 
$|R_B'| \leq 16\varepsilon n$; 
\item 
$\sum_{j=1}^{k'}\delta_G(A_j', P_i\cup B_i) + \delta_G(R_P',P_i\cup B_i)\leq
\dfrac{2C''_{\ref{thm:main-partial-part}}(\delta)}{\varepsilon}\;\lambda_k\log k \cdot d |P_i|$, for all $i\in [k']$.
\end{enumerate}
\end{theorem}


\noindent{\textbf{Remark: } We will assume that $\varepsilon\leq \delta$. If that is not the case, we can replace $\varepsilon$ with $\varepsilon' = \delta$ and hide the additional factor of $\varepsilon/\varepsilon'$ in the bound on $\phi_G(P_i\parallel B_i)$ and $\sum_{j=1}^{k'}\delta_G(A_j', P_i\cup B_i) + \delta_G(R_P',P_i\cup B_i)$ in the constant 
$C''_{\ref{thm:main-partial-part}}(\delta)$. We will also assume that $\delta \geq 1/(3k)$: indeed if $\delta <1/(3k)$, we can increase it to $1/(3k)$ and we will still have $k'\geq \lceil (1 - 2/(3k)) k \rceil = k$, as for the original value of $\delta$.}
\begin{proof}
Our algorithm consists of four steps. First, we embed the vertex set $V$ into a $k$ dimensional space using the standard spectral embedding (see Section~\ref{sec:spectral} for details). We denote the image of vertex $u$ by $\bar u$. We also let 
$\psi(\bar u) = \bar u/\|\bar u\|$ (that is, 
$\psi(\bar u)$ is the normalized  $\bar u$)
and $\mu(u) = \|\bar u\|^2$ (note: $\bar u\neq 0$ by Claim~\ref{cl:spectral-embed-norm}).
At the second step, we obtain a \emph{crude} partial partitioning $\widetilde P_1,\dots,\widetilde P_{k''}$ with buffers $\widetilde B_1,\dots,\widetilde B_{k''}$  using a new technical tool, which we introduced in Section~\ref{sec:orth-sep}. We call this tool \emph{orthogonal separators with buffers} (see Theorem~\ref{thm:orth-sep-two-buffers}). Finally, we refine the crude partitioning at the third step and discard some sets at the fourth step. 
We get subsets $P_{i},B_{i},A_{i}',A_{i}''$ for $i \in [k']$ and two extra subsets $R_P', R_B'$. We provide the pseudocode for Steps 1, 2, 3 and 4 in Figures~\ref{fig:step-1},
\ref{fig:step-2},
\ref{fig:step-3}, and
\ref{fig:step-4}.
We now analyze our algorithm.

Before we proceed to the proof, we set some notation. Let $\Ball(u,R)$ be the ball of radius $R$ around $u$ in the metric 
$\rho(u,v) = \|\psi(\bar u) - \psi(\bar v)\|$:
$$\Ball(u,R) =\{v\in V: \|\psi(\bar u)-\psi(\bar v)\|\leq R\}.$$
We define measure $\mu$ on $V$ as follows:
for every $S\subseteq V$,
$$\mu(S) = \sum_{u\in S} \mu(u).$$

\medskip \noindent{\textbf{Step 1: Spectral Embedding.}} In Section~\ref{sec:spectral}, we remind the reader the standard definition of a  spectral embedding of $G$ into $\bbR^k$. We then prove two claims about this embedding. First, we note that $\mu(V) = k$. This is a known fact (see e.g., ~\cite{LouisRTV12}). Then, in Lemma~\ref{lem:all-good}, we show that for $R < 1/\sqrt{2}$, for any vertex $u \in V$,
\begin{equation}\label{eq:ball-measure-step1}
\mu(\Ball(u,R)) \leq \frac{1}{1-2R^2}.
\end{equation}
We will use this bound with $R=\sqrt{\nicefrac{\delta}{6}}$.

\medskip \noindent{\textbf{Step 2: Crude Partial Partitioning.}} We now analyze the second step of the algorithm described in Figure~\ref{fig:step-2}. Let  $\{(\widetilde P_t,\widetilde B_t)\}_{t=1}^{T}$ be the crude partial partitioning obtained at this step. Define function
\begin{equation}
\label{def:eq:eta-uv}
\eta(u,v) = 
\begin{cases}
\|\bar u\|^2,&\text{if } u\in \widetilde P_t, v\notin \widetilde P_t\cup \widetilde B_t
\text{ for some } t;\\
\nicefrac{1}{\varepsilon}\;\|\bar u -\bar v\|^2,&\text{if } u \in \widetilde P_t,\;v\in \widetilde P_t\cup
\widetilde B_t \text{ for some } t;\\
0,&\text{otherwise.}
\end{cases}
\end{equation}
Later, we will use the following sum as an estimate of the size of the edge boundary of set $P_t$:
\begin{equation}
\label{def:eq:eta-P}
\eta(\widetilde P_t) = 
\sum_{\substack{u\in \widetilde P_t;\; v\in V;\\
s.t. (u,v)\in E}}
\eta(u,v).
\end{equation}
Note that function $\eta(u,v)$ is not symmetric. If $u$ and $v$ are in $\widetilde P_t$, then the sum above includes both 
terms $\eta(u,v)$ and $\eta(v,u)$. Depending on the argument, we will use $\eta$ to denote the cost of an edge as in 
Equation~(\ref{def:eq:eta-uv})
or the cost of all edges incident on vertices in $\widetilde P_t$ as in 
Equation~(\ref{def:eq:eta-P}).

Note that sets $\widetilde P_t, \widetilde B_t$ are contained in $X_t \cup Y_t\cup Z_t \setminus \Sigma_{t-1}$, where $X_t,Y_t,Z_t$ are orthogonal separator and its two buffers and $\Sigma_{t-1}$ are vertices covered by previous $\widetilde P_i$ for $i < t$. We define another cost function as follows:
\begin{equation}
\label{def:eq:eta-uv-2}
\tilde\eta(u,v) = 
\begin{cases}
\|\bar u\|^2,&\text{if } u\in \widetilde P_t \cup \widetilde B_t, v\notin (X_t\cup Y_t \cup Z_t)\setminus \Sigma_{t-1}
\text{ for some } i;\\
0,&\text{otherwise.}
\end{cases}
\end{equation}
We will use this cost function to bound the total cost of edges from each part in the partial partitioning $P_i$ and $B_i$ to the uncovered part $R_B'$ and $R_P'$.
The cost of all edges incident on vertices in $\widetilde P_t \cup \widetilde B_t$ for function $\tilde \eta$ is denoted as 
\begin{equation}
\label{def:eq:eta-P-2}
\tilde\eta(\widetilde P_t \cup \widetilde B_t) = 
\sum_{\substack{u\in \widetilde P_t\cup \widetilde B_t;\; v\in V;\\
s.t. (u,v)\in E}}
\tilde \eta(u,v).
\end{equation}

We prove the following lemma for all sets generated after Step 2.
\begin{lemma}\label{lem:alg-step-2}
The crude partial partitioning $\{(\widetilde P_t,\widetilde B_t)\}_{t=1}^{T}$ and subsets $R_B, R_P$ obtained at Step 2 of the algorithm satisfies the following properties:
\begin{enumerate}
    \item $\mu(\widetilde P_t) \leq 1+\delta$ for all $t$;
    \item    
     $\dfrac{1}{k}\sum_{t=1}^T\E[\mu(\widetilde P_t)]\geq 1-5\delta$;
    \item $\dfrac{1}{k}\sum_{t=1}^T\E[\mu(\widetilde B_t)]\leq 4\varepsilon$;
    \item $\dfrac{1}{k}\sum_{t=1}^T\E[\eta(\widetilde P_t)]\leq \dfrac{C_{\delta}}{\varepsilon}\cdot \lambda_k d \log k$;
    \item 
    $\sum_{t=1}^T \E|\widetilde B_t| + \E|R_B| \leq 4\varepsilon n$;
    \item 
    $\dfrac{1}{k}\sum_{t=1}^T\E[\tilde\eta(\widetilde P_t \cup \widetilde B_t)]\leq \dfrac{C_{\delta}}{\varepsilon}\cdot \lambda_k d \log k$.
\end{enumerate}
Here, the expectation is taken over the random decisions made by the algorithm at Step 2 (all other steps of the algorithm are deterministic).
\end{lemma}
\begin{proof}
We will use Theorem~\ref{thm:orth-sep-two-buffers} to analyze Step~2 of the algorithm. 
We first show item (1). Observe that
$\widetilde P_t \subset X_t$ and for every $u\in X_t$,
$X_t = \Ball(u,R) \cup (X_t\setminus \Ball(u,R))$.
Thus, 
$$\mu(\widetilde P_t)\leq \mu(\Ball(u,R)) + \mu(X_t\setminus \Ball(u,R)).$$
By Lemma~\ref{lem:all-good} (see Equation~(\ref{eq:ball-measure-step1})), $\mu(\Ball(u,R)) \leq 1/(1-\delta/3) \leq 1+\delta/2$ for all $u$. By Theorem~\ref{thm:orth-sep-two-buffers}, 
$$\min_{u\in X_t} \mu(X_t\setminus \Ball(u,R))
\leq \frac{\delta\mu(V)}{2k} =\frac{\delta}{2}.$$
Thus,
$\mu(\widetilde P_t)\leq 1 + \delta$.

We now prove item (2). Consider a vertex $u$. Observe that if $u$ gets assigned to set $\Sigma_t$ at iteration $t$, then it remains in the set $\Sigma_{t'}$ in the future iterations $t'>t$. That is,
$\Sigma_{t} \subset \Sigma_{t+1}$. Let $\Xi_t = \bigcup_{i < t} X_i \cup Y_i \cup Z_i$. Then, similarly, we have $\Xi_{t} \subset \Xi_{t+1}$. If $u$ is not in $\Xi_t$, then at step $(t+1)$, it is assigned to $\widetilde{P}_{t+1}$ with probability at least $\alpha/2$ and to $\Xi_{t+1}\setminus \widetilde{P}_{t+1}$ with probability at most $2\varepsilon \alpha$ (see Theorem~\ref{thm:orth-sep-two-buffers}). 
Thus, 
$$\Pr\{u\in \Sigma_t\mid u\in \Xi_t \} \geq \frac{\alpha/2}{\alpha/2+2\varepsilon\alpha}
= \frac{1}{1+4\varepsilon}.$$
Also, 
$$1-(1-\alpha(1+2\varepsilon))^t \geq \Pr\{u\in \Xi_t\}
\geq 1 - (1 - \alpha/2)^t.$$
Therefore (since
$T=\roundup{\nicefrac{2}{\alpha}\ln \nicefrac{1}{\delta}}$ and
$\varepsilon<\delta<1/48$), 
\begin{equation}\label{eq:bound-Sigma-T}
\Pr\{u\in \Sigma_T\}
\geq \frac{1 - (1 - \alpha/2)^T}{1+4\varepsilon}
\geq \frac{1-\delta}{1+4\delta} \geq 1 - 5\delta.
\end{equation}
Item (2) follows from the bound above because sets
$\widetilde P_t$ are disjoint and $\Sigma_T = \bigcup_{t=1}^T \widetilde P_t$.


We then prove items (3) and (5). 
Note that the remaining parts $R_P = V \setminus \Xi_T$ and $R_B = V \setminus (R_P \cup \Sigma_T \cup \Gamma_T) = \Xi_T \setminus (\Sigma_T \cup \Gamma_T)$.
Since all sets $\widetilde{B}_t$ are disjoint and $\Gamma_T = \cup_{t=1}^T \widetilde B_t$, we upper bound probabilities $\Pr\{u\in \Gamma_T\}$ and $\Pr\{u\in \Gamma_T \cup R_B\}$. Since $R_B = \Xi_T \setminus (\Sigma_T \cup \Gamma_T)$, we have $\Gamma_T \cup R_B = \Xi_T \setminus \Sigma_T$. Similar to bound~(\ref{eq:bound-Sigma-T}), we have
\begin{equation}
\Pr\{u\in \Gamma_T\} \leq \Pr\{u\in \Gamma_T \cup R_B\} \leq \Pr\{u\in \Xi_T \setminus \Sigma_T\}
\leq \frac{4\varepsilon}{1+4\varepsilon}\cdot \big(1 - (1 - \alpha(1+2\varepsilon))^T\big)\leq 
4\varepsilon,
\end{equation}
where the last inequality is due to $\Pr\{u \in \Xi_T \setminus \Sigma_T \mid u \in \Xi_T\} \leq 4\varepsilon/(1+4\varepsilon)$ and $\Pr\{u \in \Xi_T\} \leq 1-(1-\alpha(1+2\varepsilon))^T$.
Then, item (3) follows from $\Pr\{u\in \Gamma_T\} \leq 4\varepsilon$ and item (5) follows from $\Pr\{u\in \Gamma_T \cup R_B\} \leq 4\varepsilon$.

We now prove the item (4). Consider an edge $(u,v)$. We bound the probability of the event $\big\{\eta(u,v) = \|\bar u\|^2\big\}$. If $\eta(u,v) = \|\bar u\|^2$, then $u\in \widetilde P_t$,
and $v\notin \widetilde{P}_{t}\cup \widetilde{B}_{t}$ for some $t$. We first assume that $v\notin \Sigma_{t'} \cup \Gamma_{t'}$ with $t' \leq t-1$ or, in other words,
$v\notin \Sigma_{t-1} \cup \Gamma_{t-1}$. 
Then, $u\in X_t\setminus \Xi_{t-1}$ and
$v\notin X_t\cup Y_t$ for some $t$ (otherwise, if $v$ was in 
$(X_t\cup Y_t)\setminus \Sigma_{t-1} \cup \Gamma_{t-1}$, $v$ would also be in $\widetilde P_t$ or $\widetilde B_t$).
If $v\in \widetilde P_{t'} \cup \widetilde B_{t'}$ and $u\in \widetilde P_t$ with $t'<t$, then 
$v\in (X_{t'} \cup Y_{t'}) \setminus (\Sigma_{t'-1} \cup \Gamma_{t'-1})$ and
$u \notin X_{t'}\cup Y_{t'} \cup Z_{t'}$ for some $t'$.
Write,
\begin{align}
\Pr\big\{\eta(u,v) = 
\|\bar u\|^2 \big\}
&\leq 
\underbrace{
\sum_{t=1}^T\Pr\big\{u \in X_t\setminus \Xi_{t-1} \text{ and } v\notin X_t\cup Y_t\}}_{(*)} \\
&\phantom{=}+
\underbrace{
\sum_{t=1}^T\Pr\big\{v \in  (X_{t} \cup Y_{t}) \setminus (\Sigma_{t-1} \cup \Gamma_{t-1}) \text{ and } u\notin X_t\cup Y_t \cup Z_t \}}_{(**)}.
\label{eq:prob-eta-uv-bound}
\end{align}
We upper bound the first term. Two events
$\{u \in X_t;\; v\notin X_t\cup Y_t\}$ and
$\{u \notin \Xi_{t-1}\}$ are independent for every $t$. Thus,
\begin{align*}
(*)&\leq \sum_{t=1}^T\Pr\big\{u \in X_t \text{ and } v\notin X_t\cup Y_t\}
\cdot
\Pr\{u \notin \Xi_{t-1}\}\\
& =
\sum_{t=1}^T\Pr\big\{v\notin X_t\cup Y_t\mid u \in X_t\}
\cdot
\Pr\{u \in X_t\}
\cdot
\Pr\{u \notin \Xi_{t-1}\}\\
&=
\sum_{t=1}^T\Pr\big\{v\notin X_t\cup Y_t\mid u \in X_t\}
\cdot
\Pr\{u \in X_t\setminus \Xi_{t-1}\}.
\end{align*}
By Theorem~\ref{thm:orth-sep-two-buffers},
$$\Pr\{v\notin X_t\cup Y_t\mid u \in X_t \} \leq \calD\, \|\psi(\bar u) - \psi(\bar v)\|^2,$$
where $\calD=O(\nicefrac{1}{\varepsilon}\,\log \nicefrac{k}{\delta}) = O_{\delta}(\nicefrac{1}{\varepsilon}\,\log k)$. 
Observe that events
$\{u \in X_t\setminus \Xi_{t-1}\}$ 
for $t\in\{1,\dots,T\}$ are mutually exclusive. Thus,
$$
(*)
\leq 
\calD\, \|\psi(\bar u) - \psi(\bar v)\|^2
\cdot
\underbrace{\sum_{t=1}^T\Pr\{u \in X_t\setminus \Xi_{t-1}\}}_{\leq 1}\\
\leq \calD\, \|\psi(\bar u) - \psi(\bar v)\|^2.
$$
The same bound holds for $(**)$ in Equation~(\ref{eq:prob-eta-uv-bound}).
We now bound
$\E[\eta(u,v)]$:
\begin{align*}
\E\big[\eta(u,v)\big] &= \Pr\Big\{\eta(u,v) = \|\bar u\|^2 \Big\}\cdot
\|\bar u\|^2
+
\Pr\Big\{\eta(u,v) = \nicefrac{1}{\varepsilon}\|\bar u-\bar v\|^2 \Big\}\cdot
\nicefrac{1}{\varepsilon}\|\bar u - \bar v\|^2
\\
&\leq 2\calD \,\|\psi(\bar u) -\bar \psi(\bar v)\|^2\cdot \|\bar u\|^2 + \nicefrac{1}{\varepsilon}\|\bar u - \bar v\|^2.
\end{align*}
By Claim~\ref{cl:product-psi} (see below),
$\E\big[\eta(u,v)]\leq 8\calD \|\bar u-\bar v\|^2 + \nicefrac{1}{\varepsilon} \|\bar u - \bar v\|^2 = O_{\delta}(\nicefrac{1}{\varepsilon}\log k)\,\|\bar u-\bar v\|^2$.
\begin{claim}\label{cl:product-psi}
Consider two vertices $u,v\in V$ and the corresponding nonzero vectors $\bar u, \bar v$. We have
$$
\|\bar u\|^2\cdot 
\|\psi(\bar u)-\psi(\bar v)\|^2
\leq 4 \|\bar u - \bar v\|^2.
$$
\end{claim}
\noindent\textbf{Remark:} This is a known inequality. See e.g.,~\cite{CMM2} and \cite{LeeOvT12}.
\begin{proof}
Write,
$$
\|\bar u\|^2\cdot 
\|\psi(\bar u)-\psi(\bar v)\|^2
=
\|\bar u\|^2\cdot
\Big\|
\frac{\bar u}{\|\bar u\|}-\frac{\bar v}{\|\bar v\|}
\Big\|^2
=
\Big\|
\bar u -\frac{\|\bar u\|}{\|\bar v\|}
\;\bar v
\Big\|^2.
$$
We now use the \emph{relaxed triangle inequality} for squared Euclidean distance
$\|x-z\|^2 \leq 
2\|x-y\|^2 + 2\|y-z\|^2$.
We have
$$
\|\bar u\|^2\cdot 
\|\psi(\bar u)-\psi(\bar v)\|^2
\leq
2\|\bar u -\bar v\|^2
+
2\Big\|
\bar v -\frac{\|\bar u\|}{\|\bar v\;\|}
\;\bar v
\Big\|^2
\leq 4\|\bar u -\bar v\|^2.
$$
Here, we used that $\bar v$ and 
$\frac{\|\bar u\|}{\|\bar v\;\|}
\;\bar v$ are collinear vectors and, thus,
$$\Big\|\frac{\|\bar u\|}{\|\bar v\;\|}
\;\bar v - \bar v\Big\| = 
\left|\Big\|\frac{\|\bar u\|}{\|\bar v\;\|}
\;\bar v\Big\| - \|\bar v\|\right|= 
\big|\|\bar u\| - \|\bar v\|\big|\leq 
\|\bar u - \bar v\|.
$$
\end{proof}
We can now finish the proof of Lemma~\ref{lem:alg-step-2},
$$\dfrac{1}{k}\sum_{t=1}^T\E[\eta(\widetilde P_t)] = \dfrac{1}{k}\sum_{(u,v)\in E}
\E[\eta(u,v)]+\E[\eta(v,u)]=
O_{\delta}(\nicefrac{1}{\varepsilon}\log k)\,
\frac{1}{k}\sum_{(u,v)\in E}\|\bar u-\bar v\|^2.
$$
By Claim~\ref{cl:main-spectral-embed}, the right hand side is upper bounded by 
$O_{\delta}(\nicefrac{1}{\varepsilon}\log k)\,
d\lambda_k$.

Finally, we prove item (6). Similar to the analysis of item (4), for any edge $(u,v)$, we bound the probability that $\tilde\eta(u,v) = \|\bar u\|^2$. If $\tilde\eta(u,v) = \|\bar u\|^2$, then we have $u \in \widetilde P_t \cup \widetilde B_t$ and $v \not\in (X_t \cup Y_t\cup Z_t)\setminus \Sigma_{t-1}$ for some $t$. We also first assume that when $u$ is contained in $\widetilde P_t \cup \widetilde B_t$, vertex $v$ is not contained in $\Sigma_{t-1}$. Then, we must have $v \notin X_t\cup Y_t\cup Z_t$. If $v$ is covered by $\widetilde P_t$ for some $t$ before $u$ is covered, then we must have $u \notin X_t \cup Y_t$ (otherwise $u$ is contained in $\widetilde P_t \cup \widetilde B_t$). Thus, we have
\begin{align*}
\Pr\{\tilde\eta(u,v) = \|\bar u\|^2\} \leq & \sum_{t=1}^T \Pr\{u \in (X_t\cup Y_t) \setminus \Xi_{t-1} \;\text{and}\; v\notin X_t\cup Y_t\cup Z_t\} \\
& + \sum_{t=1}^T \Pr\{v \in X_t \setminus \Xi_{t-1} \;\text{and}\; u\notin X_t\cup Y_t\}.
\end{align*}
By Theorem~\ref{thm:orth-sep-two-buffers}, we have $\Pr\{\tilde\eta(u,v) = \|\bar u\|^2\} \leq 2\calD\, \|\psi(\bar u) - \psi(\bar v)\|^2$. By Claim~\ref{cl:main-spectral-embed}, we get
$$
\dfrac{1}{k}\sum_{t=1}^T\E[\tilde\eta(\widetilde P_t\cup \widetilde B_t)] = \dfrac{1}{k}\sum_{(u,v)\in E}
\E[\tilde\eta(u,v)]+\E[\tilde\eta(v,u)] = O_{\delta}(\nicefrac{1}{\varepsilon}\log k)\,
d\lambda_k.
$$
\end{proof}

By item (5) in Lemma~\ref{lem:alg-step-2} and Markov's inequality, we have $|R_B| + \sum_{t=1}^T |\widetilde B_t| \leq 16\varepsilon n$ holds with probability at least $3/4$. In the following analysis, we assume this always holds.

\medskip
\noindent{\textbf{Steps 3 \& 4.}} Our algorithm (Algorithm $1$) refines the crude partial partitioning $\{\widetilde P_t,\widetilde B_t\}_{t=1}^T$ at Step 3 and obtains set tuples 
$\{(P_t,B_t,A_t',A_t'')\}_{t=1}^T$. Then, it removes some of the sets $(P_t,B_t,A_t',A_t'')$ from the partial partitioning at Step 4. In the analysis of the algorithm, it will be more convenient for us to identify those sets $(\widetilde P_t, \widetilde B_t)$ that remain in the solution first and only then find their refinements $(P_t, B_t,A_t',A_t'')$. Let
\begin{equation}
\label{eq:def:set-I}
\calI=
\Big\{i:
\widetilde P_i \neq \varnothing,\;
\mu(\widetilde B_i)\leq C_{\delta}'\;\varepsilon\mu(\widetilde P_i),\;\text{and}\; \max\{\eta(\widetilde P_i), \tilde\eta(\widetilde P_i \cup\widetilde B_i)\}\leq C_{\delta}''/\varepsilon\cdot \; \lambda_k d \log k\;\mu(\widetilde P_i)   
\Big\},
\end{equation}
where $C'_{\delta} = 192/\delta$ and 
$C''_{\delta} = 48C_{\delta}/\delta$. We will now prove that $\Pr\{|\calI|\geq (1-2\delta)|k|\}\geq \nicefrac{1}{2}$. In the next section, we show that for each $i \in \calI$, the set tuple $(P_i,B_i,A_i',A_i'')$ satisfies all constraints at Step 3 and 4. Thus, all sets $(P_i,B_i,A_i',A_i'')$ with  $i\in\calI$ remain in the solution after Step 4 and, consequently, the algorithm succeeds with probability at least $1/4$ (We assume $|R_B| + \sum_{t=1}^T |\widetilde B_t| \leq 16\varepsilon n$ at Step 2, which holds with probability at least $3/4$).

Lemma~\ref{lem:alg-step-2} gives us upper bounds on the expected values of
$k-\sum_t \mu(\widetilde P_t)$,
$\sum_{t} \mu(\widetilde B_t)$,
$\sum_{t} \eta(\widetilde P_t)$, and $\sum_{t} \tilde\eta(\widetilde P_t\cup \widetilde B_t)$. These four random variables are non-negative. 
Thus, by Markov's inequality, with probability at least $1/2$,  the following four inequalities hold simultaneously:
\begin{align*}
\dfrac{1}{k}\sum_{t=1}^T\mu(\widetilde P_t)&\geq 1-40\delta;\\
\dfrac{1}{k}\sum_{t=1}^T\mu(\widetilde B_t)&\leq
32 \varepsilon ;\\
\dfrac{1}{k}\sum_{t=1}^T\eta(\widetilde P_t)
&\leq \nicefrac{8C_{\delta}}{\varepsilon} \;\lambda_kd \log k.\\
\dfrac{1}{k}\sum_{t=1}^T\tilde\eta(\widetilde P_t \cup \widetilde B_t)
&\leq \nicefrac{8C_{\delta}}{\varepsilon} \;\lambda_kd \log k.
\end{align*}
Denote the event that all above inequalities hold by 
$\calE$. We know that $\Pr(\calE)\geq 1/2$. Let us assume that $\calE$ occurs. Since $\delta < 1/80$, we have
\begin{align*}
\sum_{t=1}^T\mu(\widetilde B_t) 
&\leq
64\varepsilon \sum_{t=1}^T\mu(\widetilde P_t);\\
\sum_{t=1}^T\eta(\widetilde P_t)
&\leq
\nicefrac{16C_{\delta}}{\varepsilon}\,\lambda_kd \log k \sum_{t=1}^T\mu(\widetilde P_t);\\
\sum_{t=1}^T\tilde\eta(\widetilde P_t \cup \widetilde B_t)
&\leq \nicefrac{16C_{\delta}}{\varepsilon} \;\lambda_kd \log k \sum_{t=1}^T\mu(\widetilde P_t).
\end{align*}
Let $w_i = \mu(\widetilde P_i)\left/\sum_{t=1}^T\mu(\widetilde P_t)\right.$.  We rewrite the inequalities above as follows:
\begin{align*}
\sum_{i=1}^T
w_i \; \frac{\mu(\widetilde B_i)}{\mu(\widetilde P_i)}
&\leq
64\varepsilon ;\\
\sum_{i=1}^T w_i
\frac{\eta(\widetilde P_i)}{\mu(\widetilde P_i)}
&\leq
\nicefrac{16C_{\delta}}{\varepsilon}\,\lambda_kd \log k;\\
\sum_{i=1}^T w_i
\frac{\tilde\eta(\widetilde P_i \cup \widetilde B_i)}{\mu(\widetilde P_i)}
&\leq
\nicefrac{16C_{\delta}}{\varepsilon}\,\lambda_kd \log k.
\end{align*}
In the expressions above, we ignore the terms with $w_i = 0$.
Note that  $\sum_i w_i = 1$.
Suppose that we pick $i$ in $\{1,\dots, T\}$ randomly with probability $w_i$. Then, the above inequalities give bounds on the expected values of $\mu(\widetilde B_i)/\mu(\widetilde P_i)$ and 
$\eta(\widetilde P_i)/\mu(\widetilde P_i)$. By Markov's inequality, 
$$
\Pr_{i\sim w}\{i\in \calI\}=
\Pr_{i\sim w}
\Big\{\mu(\widetilde B_i)\leq C_{\delta}'\;\varepsilon\mu(\widetilde P_i)\;\text{and}\; 
\max\{\eta(\widetilde P_i), \tilde\eta(\widetilde P_i \cup\widetilde B_i)\} \leq C_{\delta}''/\varepsilon\; \lambda_k d \log k\;\mu(\widetilde P_i)
\Big\}\geq 1-\delta,
$$
where $C'_{\delta} = 192/\delta$ and 
$C''_{\delta} = 48C_{\delta}/\delta$.
Therefore, $\sum_{i\in \calI} w_i\geq 1 - \delta$. We have
$$\sum_{i\in\calI}
\mu(\widetilde P_i)\geq (1-\delta) 
\sum_{i=1}^T
\mu(\widetilde P_i) \geq (1-\delta) k.
$$
We now recall that $\mu(\widetilde P_i)\leq 1 + \delta$. Consequently,
$$|\calI|\geq \frac{1-\delta}{1+\delta} k\geq (1-2\delta)k.
$$
We just showed that if event $\calE$ occurs, then $|\calI|\geq (1-2\delta)k$ and $\Pr(\calE)\geq 1/2$. Hence,
$\Pr\{|\calI|\geq (1-2\delta)k\}\geq 1/2$.

\medskip

\noindent{\textbf{Step 3: Refined Partial Partitioning.}}
At Step 3 of the algorithm, we refine the crude partitioning
obtained at Step 2. To this end, we pick a threshold $r_i\in(0,1)$ for every pair $(\widetilde P_i,\widetilde B_i)$ with $i \in \calI$. We define the refined partitioning sets to be
\begin{itemize}
    \item $P_i = \{u\in \widetilde P_i : \mu(u) \geq r_i\}$,
    \item $B_i = \{u\in \widetilde B_i : \mu(u)\geq r_i/(1+\varepsilon)\}\cup\{u\in \widetilde P_i : \mu(u)\in [r_i/(1 +\varepsilon),r_i)\}$, 
    \item $A_i' = \{u\in \widetilde P_i : \mu(u) \leq  r_i/(1+\varepsilon)^2\}$,
    \item $A_i'' = \{u\in \widetilde P_i : \mu(u)  \in (r_i/(1+\varepsilon)^2, r_i/(1+\varepsilon))\}$.
\end{itemize}
The threshold $r_i$ must satisfy five  conditions: (1) $|B_i|\leq C'_{\ref{thm:main-partial-part}}(\delta)\varepsilon |P_i|$; (2) $\phi_G(P_i\parallel B_i)
\leq \nicefrac{C''_{\ref{thm:main-partial-part}}(\delta)}{\varepsilon}\;\lambda_k \log k$; (3) $|A''_i| \leq 10\varepsilon |P_i|$, and (4) $\delta_G(A'_i,P_i\cup B_i) \leq \nicefrac{C''_{\ref{thm:main-partial-part}}(\delta)}{\varepsilon}\;\lambda_k \log k\cdot d|P_i|$; (5) $\delta_G(P_i\cup B_i, (\Sigma_T\cup R_P) \setminus \widetilde P_i) \leq \nicefrac{C''_{\ref{thm:main-partial-part}}(\delta)}{\varepsilon}\;\lambda_k \log k\cdot d|P_i|$. 
At Step 4, we drop sets $(P_i,B_i,A_i',A_i'')$ for which we could not find such threshold. We now show that for every $i\in\calI$ such threshold $r_i$ exists (set $\calI$ is defined in Equation~(\ref{eq:def:set-I})). We use the probabilistic method. 
\begin{lemma}\label{lem:alg-step-3}
Consider $i\in \calI$. Suppose, we select elements in sets $P_i$ and $B_i$ using a random threshold $r_i$,
which is uniformly distributed in $(0,1)$. Then
\begin{enumerate}
\item 
$\E_{r_i}|B_i|\leq 2C'_{\delta}\,
\varepsilon
\E_{r_i}|P_i|$;
\item 
$\E_{r_i}\Big[\delta_G(P_i,V\setminus (P_i\cup B_i))\Big]\leq
\frac{C''_{\delta}}{\varepsilon}\lambda_k\log k\cdot d\E_{r_i}|P_i|$;
\item 
$\E_{r_i}|A_i''| \leq 2\varepsilon \E_{r_i}|P_i|$;
\item
$\E_{r_i}\Big[\delta_G(A_i',P_i \cup B_i) \Big] \leq 
\frac{C''_{\delta}}{\varepsilon}\lambda_k\log k \cdot d\E_{r_i}|P_i|.$
\item 
$\E_{r_i}\Big[\delta_G(P_i \cup B_i, (\Sigma_T\cup R_P) \setminus \widetilde P_i) \Big] \leq 
\frac{C''_{\delta}}{\varepsilon}\lambda_k\log k \cdot d\E_{r_i}|P_i|.$
\end{enumerate}
\end{lemma}
\begin{proof}
Denote 
$$B'_i = \{u\in \widetilde B_i : \mu(u)\geq r_i/(1+\varepsilon)\}\text{ and }
B''_i = \{u\in \widetilde P_i : \mu(u)\in [r_i/(1+\varepsilon),r_i)\}.$$
Then, $B_i = B'_i \cup B''_i$. Write,
$$\E_{r_i}|P_i| = 
\sum_{u\in \widetilde P_i }\Pr_{r_i}\{u\in P_i\} = 
\sum_{u\in \widetilde P_i } \Pr_{r_i}\{r_i \leq \mu(u)\}
= \mu(\widetilde P_i).$$
Here, we used that $\mu(u)\leq 1$ for all $u$ (see Claim~\ref{cl:spectral-embed-norm}). 
Similarly, $\E|B'_i|\leq (1+\varepsilon)\mu(\widetilde B_i)$. Then, 
$$\E|B''_i| = 
\sum_{u\in \widetilde P_i }\Pr_{r_i}
\Big\{\mu(i)\in [r_i/(1+\varepsilon),r_i]\Big\} = 
\sum_{u\in \widetilde P_i }\Pr_{r_i}
\Big\{r_i \in [\mu(i),
(1+\varepsilon)\mu(i)]\Big\} \leq \varepsilon \mu(\widetilde P_i).
$$
Thus, using the definition of set $\calI$, we get
$$
\E|B_i| \leq
\E|B'_i| + \E|B''_i|
= (1+\varepsilon)
\mu(\widetilde B_i)
+ \varepsilon \mu(\widetilde P_i)
\leq ((1+\varepsilon)C'_{\delta} + 1)
\varepsilon\mu(\widetilde P_i) = 
2C'_{\delta}\;\varepsilon
\E|P_i|.
$$
This proves the first claim of Lemma~\ref{lem:alg-step-3}. 

We assign all vertices $u \in \widetilde P_i$ with $\mu(u) \in (r_i/(1+\varepsilon)^2,r_i/(1+\varepsilon))$ to set $A''_i$.
Then, we have
\begin{multline*}
\E|A''_i| = 
\sum_{u\in \widetilde P_i }\Pr_{r_i}
\Big\{\mu(i)\in (r_i/(1+\varepsilon)^2,r_i/(1+\varepsilon))\Big\} =\\= 
\sum_{u\in \widetilde P_i }\Pr_{r_i}
\Big\{r_i \in [(1+\varepsilon)\mu(i),
(1+\varepsilon)^2\mu(i)]\Big\}
=
\sum_{u\in \widetilde P_i }(\varepsilon + \varepsilon^2)\mu(i)
<
2\varepsilon \mu(\widetilde P_i).
\end{multline*}
Since $\mu(\widetilde P_i) = \E_{r_i} |P_i|$, we get the third claim. 

To show claims 2 and 4 of Lemma~\ref{lem:alg-step-3}, we bound the expected number of edges from set $P_i$ to set $V \setminus (P_i \cup B_i)$, and the expected number of edges from set $A_i'$ to set $P_i \cup B_i$.

\begin{claim}\label{cl:uv-separated-step-3} Consider an edge $(u,v)\in E$ with $u\in \widetilde P_i$. We have
$$\Pr\{u\in P_i;\;v\notin P_i\cup B_i\}\leq 2\eta(u,v),$$
and
$$
\Pr\{u\in A_i';\;v\in P_i\cup B_i\} \leq 2\eta(u,v).
$$
\end{claim}
\begin{proof}
Consider two cases. If $v\in \widetilde P_i\cup   \widetilde B_i$, then 
\begin{align*}
\Pr\{u\in P_i,\;v\notin P_i \cup B_i \} &= 
\Pr\{\mu(u)\geq r_i
\text{ and }
\mu(v)< r_i/(1+\varepsilon)
\}\\ 
&\leq \Pr\big\{
r_i\in\big[
(1+\varepsilon)
\mu(v), \mu(u)\big]
\big\}\\
&\leq
\mu(u) - (1+\varepsilon)\mu(v)
.
\end{align*}
By Claim~\ref{cl:a-minus-b}, 
$$\mu(u) - (1+\varepsilon)\mu(v)
= \|\bar u\|^2 - 
(1+\varepsilon)\|\bar v\|^2 
\leq (1+\nicefrac{1}{\varepsilon})(\|\bar u\| - \|\bar v\|)^2 \leq 2(\|\bar u\| - \|\bar v\|)^2/\varepsilon.
$$
Using the triangle inequality $\|\bar u\|-\|\bar v\|\leq \|\bar u-\bar v\|$, we conclude that 
$$\Pr\{u\in P_i,\;v\notin P_i \cup B_i \}\leq 2\eta(u,v).$$
Similarly, we have
\begin{align*}
\Pr\{u\in A'_i \text{ and } v \in P_i \cup B_i \} &= 
\Pr\{\mu(u)\leq r_i/(1+\varepsilon)^2
\text{ and }
\mu(v) \geq r_i/(1+\varepsilon)
\}\\ 
&\leq \Pr\big\{
r_i\in\big[
(1+\varepsilon)^2
\mu(u), (1+\varepsilon)\mu(v)\big]
\big\}\\
&\leq
(1+\varepsilon)(\mu(v) - (1+\varepsilon)\mu(u)) \leq 2(\|\bar u\| - \|\bar v\|)^2/\varepsilon
.
\end{align*}
Therefore, we have 
$$
\Pr\{u\in A_i',\;v\in P_i \cup B_i \}\leq 2\eta(u,v).
$$

If $v\notin  \widetilde P_i \cup  \widetilde B_i$, then $\Pr\{u\in A_i',\;v\in \widetilde P_i \cup \widetilde B_i \} = 0$, and 
$$
\Pr\{u\in P_i,\;v\in P_i \cup B_i \} = \Pr\{u\in P_i\} = \|\bar u\|^2 = \eta(u,v).
$$
\end{proof}

By Claim~\ref{cl:uv-separated-step-3}, the expected number of edges from set $P_i$ to set $V\setminus(P_i \cup B_i)$ is at most $2\eta(\widetilde P_i)$. Also, the expected number of edges from set $A_i'$ to set $P_i \cup B_i$ is at most $2\eta(\widetilde P_i)$. In other words, $\E\big[\delta_G(P_i,V\setminus (P_i\cup B_i))\big]\leq 2\eta(\widetilde P_i)$ and $\E\big[\delta_G(A_i',P_i \cup B_i\big]\leq 2\eta(\widetilde P_i)$. Using the definition of set $\calI$ (see~(\ref{eq:def:set-I})), we get the claims 2 and 4 of Lemma~\ref{lem:alg-step-3}.

Finally, we prove claim 5 of Lemma~\ref{lem:alg-step-3}. We have for any edge $(u,v)$ with $u \in \widetilde P_i\cup \widetilde B_i$, 
$$
\Pr\{u \in P_i\cup B_i, v \in (\Sigma_T \cup R_P)\setminus \widetilde P_i \} \leq \Pr\{u \in P_i\cup B_i\} = (1+\varepsilon)\|\bar u\|^2 \leq 2\tilde\eta(u,v).
$$
Thus, the expected number of edges from $P_i\cup B_i$ to $(\Sigma_T \cup R_P)\setminus \widetilde P_i$ is at most $2\tilde \eta(\widetilde P_i\cup \widetilde B_i)$. By the definition of set $\calI$ (see~(\ref{eq:def:set-I})), we get the conclusion.
\end{proof}

Using Lemma~\ref{lem:both-hold} with six random variables, we conclude that there exists $r_i \in (0,1)$ such that inequalities (1) $|B_{i}|\leq 10C_\delta' \;\varepsilon |P_i|$, (2) $\delta_G(P_i,V\setminus(P_i\cup B_i))\leq \nicefrac{5C''_{\delta}}{\varepsilon}\;\lambda_k \log k \cdot d|P_i|$, (3) $|A''_i| \leq 10\varepsilon |P_i|$, (4) $\delta_G(A'_i,P_i\cup B_i) \leq \nicefrac{5C''_{\delta}}{\varepsilon}\;\lambda_k \log k\cdot d|P_i|$, and (5) $\delta_G(P_i\cup B_i,(\Sigma_T\cup R_P)\setminus \widetilde P_i) \leq \nicefrac{5C''_{\delta}}{\varepsilon}\;\lambda_k \log k\cdot d|P_i|$ hold simultaneously.
The second inequality is equivalent to 
$\phi(P_i\parallel B_i) \leq
\nicefrac{5C''_{\delta}}{\varepsilon}\;\lambda_k \log k$. 
In this theorem, we use the following functions $C'_{\ref{thm:main-partial-part}}$ and $C''_{\ref{thm:main-partial-part}}$: 
$C'_{\ref{thm:main-partial-part}}(\delta)=10C'_{\delta}$ and 
$C''_{\ref{thm:main-partial-part}}(\delta) = 5C''_{\delta}$.
Combining the inequalities (4) and (5), we get the property (6) in Theorem~\ref{thm:main-partial-part}.
In Algorithm 1, all sets $P_i,B_i,A_i',A_i''$ for $i \in [k']$ and $R_B', R_P'$ are disjoint and cover the entire graph. Since all set tuples $(P_i,B_i,A_i',A_i'')$ with $P_i = \varnothing$ are discarded at Step 4, all sets $P_i$ returned by Algorithm 1 are nonempty. Note that $R_B' \subseteq R_B \cup \bigcup_{i=1}^T \widetilde B_i$. Since we assume $|R_B| + \sum_{i=1}^T |\widetilde B_i| \leq 16 \varepsilon n$ at Step 2 (This condition holds with probability at least $3/4$), we have $|R_B'| \leq 16\varepsilon n$. 
This finishes the proof of Theorem~\ref{thm:main-partial-part}.
\end{proof} 

\section{From Disjoint Sets to Partitioning}\label{sec:partial-tocomplete}

We now show how to use the partial partitioning given by Algorithm 1 in Section~\ref{sec:partial-partition}  to obtain a true $\varepsilon$-buffered partitioning. 
We prove the following lemma.

\begin{lemma}\label{lem:partial-to-complete}
Consider a $d$-regular graph $G$. 
Let $\{(P_i,B_i, A_i',A_i'')\}_{i\in [k']}$ and $R_P',R_B'$ be a partial $\varepsilon$-buffered partitioning of $G$ given by Algorithm 1. 
Then, for every $k \in \{1,\cdots,k'\}$ and $\delta' = (k'- k + 1)/k'$, we can convert this partial partitioning into a true $\nicefrac{54\varepsilon}{\delta'}$-buffered partitioning $P'_1,\dots, P'_{k}$, $B'_1,\dots, B'_{k}$ of $G$ such that
$$\phi_G(P'_1,\dots,P'_{k}\parallel B'_1,\dots, B'_{k})\leq
\frac{4C''_{\ref{thm:main-partial-part}}(\delta)}{\delta'} \cdot\frac{\log k}{\varepsilon}\lambda_k.
$$
\end{lemma}

\begin{proof}
Let us sort all pairs $(P_i,B_i,A_i',A_i'')$ by size and assume $|P_1| \leq \cdots \leq |P_{k'}|$.
Now, we generate the true buffered partitioning of the graph. 
The true buffered partitioning $(P'_i,B'_i)$ contains the pairs $(P_i,B_i)$ for $i \in [k-1]$ in the partial partitioning and a pair of new sets  $(P'_{k}, B'_{k})$. Specifically, we let  
$P'_i=P_i$ and $B'_i = B_i$ for $i\in [k-1]$ and 
$$
P'_{k} = R_P'\cup \bigcup_{j=1}^{k'} A_j' \cup \bigcup_{j=k}^{k'} P_j;
\;\;\;\;\;\;
B'_{k} = R_B' \cup \bigcup_{j=1}^{k'} A_j'' \cup \bigcup_{j=k}^{k'} B_j. 
$$
We can think of each set $A_i''$ is the buffer for the set $A_i'$ for $i \in [k']$, and the set $R_B'$ is the buffer for the set $R_P'$. We also combine these sets and buffers with the largest $k'-k+1$ pairs $(P_i,B_i)$ for $i = k,k+1,\cdots,k'$ in the partial partitioning, respectively.   

By Theorem~\ref{thm:main-partial-part}, all sets $P_i,B_i, A_i',A_i''$ and $R_P',R_B'$ are disjoint and cover the entire graph. Also, all sets $P_i$ and $R_P'$ are nonempty. Thus, all sets $P'_i$ are disjoint and nonempty, and $\bigcup_{i=1}^{k}  P'_i\cup B'_i = V$. Also, for all $i \in [k-1]$, we have 
$|B_i|\leq \varepsilon |P_i|$ and
\begin{equation}\label{eq:p-prime-vs-max}
\phi_G(P'_i,B'_i)= \phi_G(P_i,B_i) \leq \dfrac{C''_{\ref{thm:main-partial-part}}(\delta)}{\varepsilon}\;\lambda_k \log k.
\end{equation} 

It remains to verify that the last pair of sets $P'_{k}$ and $B'_{k}$ satisfy the required conditions. By items 4 and 5 of Theorem~\ref{thm:main-partial-part}, we have  
$$
|B'_k| \leq |R'_B| + \sum_{j=1}^{k'} |A_j''| +\sum_{j=1}^{k'} |B_j| \leq 16\varepsilon n + 11\varepsilon \sum_{j=1}^{k'} |P_i| \leq 27\varepsilon n.
$$
Since $|P_1| \leq \cdots \leq |P_{k'}|$, we have $\sum_{i=1}^{k-1} |P_i| \leq \nicefrac{k-1}{k'} \sum_{i=1}^{k'} |P_i|$. Thus, we have 
\begin{multline*}
    |P'_k| = |V| - |R_B'| - \sum_{i=1}^{k'} |A_i''|+|B_i| -\sum_{i=1}^{k-1} |P_i|  \geq \\
    \geq \left(1-\frac{k-1}{k'}\right) \cdot \left(|V| - |R_B'| - \sum_{i=1}^{k'} |A_i''|+|B_i|\right)\geq \delta' (n-27\varepsilon n) \geq \delta' n/2.
\end{multline*}
Hence, we have $|B'_k| \leq \nicefrac{54\varepsilon}{\delta'} |P'_k|$. 

We now bound the buffered expansion of this last part. By items (3) and (6) of Theorem~\ref{thm:main-partial-part}, we have
\begin{align*}
\phi_G(P'_k \parallel B'_k)
&\leq 
\frac{\sum_{i=1}^{k-1} \delta_G (P'_k, P_i\cup B_i)}{d|P'_k|}\\
&\leq \frac{\sum_{i=1}^{k-1}\sum_{j=1}^{k'} \delta_G (A_j', P_i\cup B_i) + \delta_G(R_P', P_i\cup B_i) + \sum_{j=k}^{k'} \delta_G(P_j, P_i\cup B_i)}{d \cdot \delta' n/2}\\
&\leq \frac{2C''_{\ref{thm:main-partial-part}}(\delta)/\varepsilon \cdot \lambda_k \log k \cdot d\sum_{i=1}^{k-1}|P_i| + \sum_{j=k}^{k'} \delta_G(P_j, V\setminus(P_j\cup B_j))}{d \cdot \delta' n/2}\\
&\leq
\frac{4C''_{\ref{thm:main-partial-part}}(\delta) / \delta'}{\varepsilon}\cdot \lambda_k\log k
.
\end{align*}
This concludes the proof of Lemma~\ref{lem:partial-to-complete}.
\end{proof}

We now prove the main result of the paper,  Theorem~\ref{thm:main}.
\begin{proof}[Proof of Theorem~\ref{thm:main}]
Let $\hat{k} = \lfloor(1+\delta)k\rfloor$ and $\hat{\delta} = \min\{\nicefrac{(1-1/\sqrt{1+\delta})}{2}, 1/80\}$. Let $k' = \lceil(1-2\hat{\delta})\hat{k}\rceil$ and $\delta' = (k'-k+1)/k'$. 
We first use Algorithm 1 from Section~\ref{sec:partial-partition} with parameters $\hat{k}$, $\hat{\varepsilon} = \varepsilon \delta'/54$, and $\hat{\delta}$ to obtain a partial $\hat{\varepsilon}$-buffered partitioning $(P_1,B_1, A_1',A_1''),\dots,(P_{k'},B_{k'}, A_{k'}',A_{k'}'')$.
By Theorem~\ref{thm:main-partial-part}, the buffered expansion of each set $P_i$ with buffer set $B_i$ is at most $\nicefrac{C''_{\ref{thm:main-partial-part}}(\delta)}{\hat{\varepsilon}}\;\lambda_{\hat{k}} \log \hat{k}$. 
Then, we apply Lemma~\ref{lem:partial-to-complete} to transform this partial partitioning into a true $k$ partitioning. 
Since $k' = \lceil(1-2\hat{\delta})\hat{k}\rceil$, we have $k' \geq \sqrt{1+\delta}k-1$. Then, we have $\delta' \geq 1-1/\sqrt{1+\delta}$. By Lemma~\ref{lem:partial-to-complete},  the expansion of this $\varepsilon$-buffered $k$ partitioning is at most $\nicefrac{c(\delta)}{\varepsilon}\;\lambda_{\hat{k}} \log \hat{k}$, where $c(\delta) = \nicefrac{4C''_{\ref{thm:main-partial-part}}(\delta)}{\delta'}$ is a function that only depends on $\delta$.
\end{proof}

\section{Spectral Embedding}\label{sec:spectral}
Consider a $d$-regular graph $G$. Let $L_G$ be its normalized Laplacian. Let $x_1,\dots,x_n$ be an orthonormal eigenbasis for $L_G$ and $\lambda_i$ be the eigenvalue of $x_i$. Without loss of generality, we assume that $\lambda_1\leq \dots\leq \lambda_n$. Note that $\lambda_1 = 0$, so we may assume that $x_1=\mathbf{1}/\sqrt{n}$. Define an $k\times n$ matrix $U = (x_1,\dots,x_k)^T$;
that is, the $(i,u)$ entry of $U$ equals $U(i,u) = x_i(u)$ where $i\in [k]$ and $u\in V$. 
Rows of $U$ are indexed by integers from 1 to $k$ and columns by  vertices $u\in V$ of the graph (to simplify notation, we may assume that $V = [n]$).
Note that $UU^T = I_k$, since vectors $x_1,\dots, x_k$ are orthonormal.
Let $\{e_u\}_{u\in V}$ be the standard orthonormal basis in ${\mathbb R}^V$.


We are ready to define the spectral embedding of $G$. Let $\bar u$ be the column of $U$ indexed by vertex $u$.
The spectral embedding maps vertex $u$ to vector $\bar u$.

Define $\psi(u) = u_i / \|u_i\|$. For a subset of vertices $S \subseteq V$, let $\mu(S) = \sum_{u\in S}\|\bar u\|^2$ be the measure of set $S$.
Now we will state and prove basic properties of the spectral embedding.

\begin{claim}\label{cl:spectral-embed-norm}
    For all $u \in V$, we have $0 < \|\bar u\| \leq 1$.
\end{claim}
\begin{proof}
    Since $x_1 = \mathbf{1}/\sqrt{n}$, for all $u \in V$, we have $\bar u(1) = 1/\sqrt{n}$ and $\|\bar u\| \geq 1/\sqrt{n} > 0$.
Further,
    $$
    \|\bar u\|^2 = \sum_{i=1}^k \bar u(i)^2 = \sum_{i=1}^k x_i(u)^2 =
    \sum_{i=1}^k \langle x_i, e_u \rangle^2 
    \leq \sum_{i=1}^n \langle x_i,e_u \rangle^2 = \|e_u\|^2 = 1.
    $$
\end{proof}

\begin{claim}\label{cl:main-spectral-embed}
We have
\begin{enumerate}
    \item $\sum_{u\in V} \|\bar u\|^2 = k$
    \item $\sum_{(u,v)\in E} \|\bar u - \bar v\|^2 \leq kd \lambda_k$
\end{enumerate}
\end{claim}
\begin{proof}
Note that the $(u,v)$ entry of matrix $U^TU$ equals $\langle \bar u, \bar v\rangle$, since $U$ has columns $\bar u$ for $u\in V$.

\medskip\noindent\textit{1.} We have,
$\sum_{u\in V} \|\bar u\|^2 = \tr(U^TU) = \tr(UU^T) = \tr I_k = k$, as required.

\medskip

\noindent \textit{2.}
We have,
\begin{align*}\sum_{(u,v)\in E} \|\bar u - \bar v\|^2 &= 
\sum_{(u,v)\in E}\sum_{i=1}^k \|\bar u(i) - \bar v(i)\|^2
= \sum_{i=1}^k\sum_{(u,v)\in E} \|x_i(u) - x_i(v)\|^2\\
&\stackrel{\tiny\text{by (\ref{eq:lapl})}}{=}
d\sum_{i=1}^k x_i^TL_Gx_i = d \sum_{i=1}^k\lambda_i \leq dk \lambda_k,
\end{align*}
where 
we used that $\lambda_1\leq \dots \leq \lambda_k$ in the last inequality.
\end{proof}

We show that the spectral embedding vectors $\{\psi(\bar{v})\}$ satisfy the following spreading property. It is a variant of Lemma 3.2  from the paper by Lee, Oveis-Gharan and Trevisan~\cite{LeeOvT12}. 
\begin{lemma}\label{lem:all-good}
Assume that we are given a parameter $R\in [0, 1/\sqrt{2}]$. For every vertex $u$, consider the ball of radius $R$ around $u$, $\Ball(u,R) = \{v:\|\psi(\bar u) - \psi(\bar v)\| \leq R\}$. Then $\mu(\Ball(u,R)) \leq 1/(1-2R^2)$ for every $u$.
\end{lemma}
\begin{proof}
Consider a vertex $u\in V$ and $C=\Ball(u,R)$. 
Let $a_v = \|\bar v\|$ for $v\in C$. 
Then, $\bar v = a_v \psi(\bar v)$ for $v\in C$.
We have, $\mu(C) = \sum_{v\in C}a_v^2$. By the definition of $C$, $\|\psi(\bar u) - \psi(\bar v)\| \leq R$ for $v\in C$
and hence 
$\|\psi(\bar v) - \psi(\bar w)\| \leq 2R$ for all pairs $v,w\in C$. Therefore,
\begin{equation}\label{ineq:psi-product}
    \langle \psi(\bar v), \psi(\bar w)\rangle = 1 - \frac{\|\psi(\bar v) - \psi(\bar w)\|^2}{2}  \geq 1 - 2R^2  \quad\text{for all\;}  v,w\in C.
\end{equation}    
Write,
$$\mu(C) = 
\sum_{v\in C} a_v^2 = 
\frac{1}{\sum_{v\in C} a_v^2}
\sum_{v,w\in C} a_{v}^2 a_{w}^2.
$$
By inequality~(\ref{ineq:psi-product}), 
$$a_{v} a_{w}
\leq \frac{a_{v} a_{w}\langle
\psi(\bar v), \psi(\bar w) \rangle}{1-2R^2} = 
\frac{\langle
\bar v, \bar w \rangle}{1-2R^2}.
$$
Thus,
$$\mu(C) \leq 
\frac{1}{\sum_{v\in C} a_v^2}
\sum_{v,w\in C} 
\frac{a_{v} a_{w}\,\langle \bar v,\bar w\rangle}{1-2R^2}. 
$$
For any vertex $v \in V$, let $e_v \in \bbR^V$ be the standard basis vector where $e_v(v) = 1$ and $e_v(u) = 0$ for all $u\neq v$. Let 
$$z = \frac{\sum_{v\in C} a_v e_v}{\sqrt{\sum_{v\in C} a_v^2}}.$$
For any standard basis vector $e_v$, we have $Ue_v = \bar v$.
Therefore, 
$$Uz = \frac{1}{\sqrt{\sum_{v\in C} a_v^2}}
\sum_{v\in C} a_{v}\bar v,$$ 
and 
$$\mu(C) \leq \frac{z^T (U^TU)z}{(1-2R^2)}.$$

We prove that $\|Uz\|^2 = z^T(U^TU)z \leq 1$. To this end, note that $z$ is a unit vector and $\|Uz\|^2 \leq \sigma_{max}(U)^2 = \sigma_{max}(U^T)^2$, where $\sigma_{max}(U)$ and $\sigma_{max}(U^T)$ are the largest singular values of $U$ and $U^T$, respectively (here, we used the definition of singular values and the fact that matrices $U$ and $U^T$ have the same non-zero singular values). Since $UU^T= I_d$, all singular values of $U^T$ are equal to $1$. We conclude that 
$\|Uz\|^2 \leq 1$.

\end{proof}

\section{Orthogonal Separators with Buffers -- Proofs}\label{sec:os-proof}

In this section, we show the algorithm that generates orthogonal separators with buffers. We prove Theorem~\ref{thm:orth-sep}, Theorem~\ref{thm:cor:orth-sep-main}, and Theorem~\ref{thm:orth-sep-two-buffers}.

\newtheorem*{thm:os-buffer}{Theorem \ref{thm:orth-sep}}
\begin{thm:os-buffer}
There exists a randomized polynomial-time procedure that given a finite set $U$ of unit vectors in $\bbR^d$ and positive parameters $\varepsilon \in (0,1), m \geq 3, R \in (0,2)$, returns an $m$-orthogonal separator with an $\varepsilon$-buffer with distortion $\calD = O_R(\nicefrac{1}{\varepsilon}\;\log m)$, separation radius $R$, and probability scale $\alpha \geq O_R(1/poly(m))$.

For two disjoint random sets $X, Y\subset U$ chosen from this orthogonal separator distribution, we have the following properties:
\begin{enumerate}
\item For all $\bar u \in U$, $\Pr\{\bar u \in X\} = \alpha$; (for some $\alpha$ that depends on $m$ and $R$).
\item For all $\bar u \in U$, $\Pr\{\bar u \in Y\} \leq \varepsilon\alpha$.
\item For all $\bar u, \bar v \in U$ with 
$\|\bar u - \bar v\| \geq R$,
$\Pr\{\bar v \in X\ \given \bar u \in X\} \leq 
\frac{1}{m}
$.
\item For all $\bar u, \bar v \in U$,
$\Pr\{\bar v\notin X\cup Y \mid \bar u \in X\} \leq \calD\, \|\bar u - \bar v\|^2$, where $\calD = O_R(\nicefrac{1}{\varepsilon}\;\log m)$.
\end{enumerate}
\end{thm:os-buffer}



\begin{proof}[Proof of Theorem~\ref{thm:orth-sep}]
We use the following procedure to generate orthogonal separators with buffers.
We sample a $d$-dimensional Gaussian vector $g \sim \calN(0,I_d)$. For every vector $\bar{u}$ in $U$, we let $g_u = \langle\bar{u},g\rangle$ be the projection of vector $\bar{u}$ on the direction $g$. For a standard gaussian random variable $Z\sim \calN(0,1)$, we use $\FFF(t) = \Pr\{Z \geq t\}$ to denote the probability that $Z \geq t$.
We pick a threshold $t$ such that $\FFF(t) = \alpha$ for some $\alpha$ that we will specify later; our choice of $\alpha$ will guarantee that $t\leq 1$. Let $\varepsilon' = \varepsilon/ (e(t+1/t))$.  Then, we construct the orthogonal separator $X$ and the buffer $Y$ as follows:
$$
X = \{\bar{u}: g_u \geq t\}; \qquad Y = \{\bar{u}: t-\varepsilon' < g_u < t \}.
$$
Now we show that this procedure satisfies the required properties. 

1. For every vector $\bar{u} \in U$, we have 
$$
\Pr\{\bar{u} \in X\} = \Pr\{g_u \geq t\} = \FFF(t) = \alpha.
$$

2. For every vector $\bar{u} \in U$, we have 
\begin{multline*}
\Pr\{\bar{u} \in Y\} = \Pr\{t-\varepsilon' < g_u < t\} \leq 
\nc e^{-\frac{(t-\varepsilon')^2}{2}}\cdot \varepsilon' 
\leq\\ \leq
\frac{\varepsilon' e^{\varepsilon' t}}{\sqrt{2\pi}} e^{-\frac{t^2}{2}} =
\frac{\varepsilon e^{\varepsilon}}{e\sqrt{2\pi}(t+1/t)} e^{-\frac{t^2}{2}} \leq
\frac{e^{\varepsilon}}{e}\cdot 
\varepsilon\FFF(t) \leq \varepsilon \alpha,
\end{multline*}
where the third inequality is due to Lemma~\ref{lem:NormalDistrib}.

3. For every $\bar{u},\bar{v} \in U$, we have
$$
\Pr\{\bar{u} \in X, \bar{v} \in X\} = \Pr\{g_u \geq t, g_v \geq t\} \leq \Pr \{(g_u+g_v)/2 \geq t \}.
$$
We know that $g_u, g_v$ are both random Gaussian variables from $\calN(0,1)$. Thus, we have $(g_u+g_v)/2$ is also a Gaussian variable with variance 
$$
\Var\left[\frac{g_u + g_v}{2}\right] = \frac{1}{4}\E[(g_u+g_v)^2] = \frac{1}{4}(2+2\langle \bar{u},\bar{v}\rangle) = 1-\frac{\|\bar{u}-\bar{v}\|^2}{4},
$$
where the second equality is due to $\E[g_ug_v] = \langle\bar{u},\bar{v}\rangle$ and the third equality used $\bar{u},\bar{v}$ are unit vectors.
Thus for every $\bar{u},\bar{v} \in U$ with $\|\bar{u}-\bar{v}\| \geq R$, we have $\Var\left[\sfrac{(g_u + g_v)}{2}\right] \leq 1-\sfrac{R^2}{4}$. From Lemma~\ref{lem:Gaussian_1} we get that there exists a constant $C$ such that
$$
\Pr \left\{\frac{g_u+g_v}{2} \geq t \right\} \leq \FFF\left(\frac{t}{\sqrt{1-R^2/4}} \right) \leq \frac{1}{t}(Ct\FFF(t))^{\frac{1}{\sqrt{1-R^2/4}}}.
$$
Since $\FFF(t) = \alpha$, we have
$$
\Pr\{\bar{u} \in X, \bar{v} \in X\} \leq \Pr \left\{\frac{g_u+g_v}{2} \geq t \right\} \leq \alpha \cdot C(C t\alpha)^{\frac{1}{\sqrt{1-R^2/4}}-1}. 
$$
By Lemma~\ref{lem:NormalDistrib}, we have $t = \Theta(\sqrt{\log 1/\alpha})$. Then we can find some $\alpha \geq 1/poly(m)$ (for a fixed $R$) that depends on $m$ and $R$ such that $\Pr\{\bar{u} \in X, \bar{v} \in X\} \leq \alpha/m$. 
Since $\Pr\{\bar{u} \in X\} = \alpha$, we have 
$$
\Pr\{\bar{v} \in X \given \bar{u} \in X\} 
\leq \frac{1}{m}.
$$

4. For every $\bar{u},\bar{v} \in U$, we have
$$
\Pr\{\bar{u} \in X , \bar{v} \notin X \cup Y \} = \Pr\{g_u \geq t, g_v \leq t-\varepsilon' \}.
$$
Since $g$ is a standard Gaussian random vector, we have $g_u$ and $g_v$ are jointly Gaussian random variables with distribution $\calN(0,1)$. Since $\varepsilon \leq 1$ and $t = \Theta(\sqrt{\log m})$, we have $\varepsilon' = \varepsilon/(e(t+1/t)) < t$. Using Lemma~\ref{lem:separate} on $g_u,g_v$ with parameters $\hat{m} = 1/\alpha$ and $\hat{\varepsilon} = \varepsilon'$, we get
$$
\Pr\{g_u \geq t, g_v \leq t-\varepsilon' \} \leq O\left(\frac{\sqrt{\log \hat{m}}}{\varepsilon' \hat{m}}\right) \cdot\|\bar{u}-\bar{v}\|^2 \leq \alpha \calD \|\bar{u}-\bar{v}\|^2,
$$
where $\calD = O_R(\nicefrac{1}{\varepsilon}\log m)$.
\end{proof}

\newtheorem*{thm:cor-os-buffer}{Theorem \ref{thm:cor:orth-sep-main}}
\begin{thm:cor-os-buffer}
There exists a randomized procedure that given a finite set $U$ of unit vectors in $\bbR^d$ equipped with a measure $\mu$ and positive parameters $\varepsilon \in (0,1), \delta \leq 2/3, R \in (0,2)$, returns an $\delta$-orthogonal separator with an $\varepsilon$-buffer with distortion $\calD = O_R(\nicefrac{1}{\varepsilon}\;\log \nicefrac{1}{\delta})$, separation radius $R$, and probability scale $\alpha \geq O_R(1/poly(m))$.

For two disjoint random sets $X, Y\subset U$ chosen from this orthogonal separator distribution, we have the following properties:
\begin{enumerate}
\item For all $\bar u \in U$, $\Pr\{\bar u \in X\} \in [\alpha/2, \alpha]$.
\item For all $\bar u \in U$, $\Pr\{\bar u \in Y\} \leq \varepsilon \alpha$.
\item 
$\min_{\bar u \in X} \mu(X\setminus \Ball(\bar u, R))\leq \delta \mu(U)$ (always).
\item For all $\bar u, \bar v \in U$,
$\Pr\{\bar v\notin X\cup Y \mid \bar u \in X\} \leq \calD\, \|\bar u - \bar v\|^2$, where $\calD = O_R(\nicefrac{1}{\varepsilon}\;\log \nicefrac{1}{\delta})$.
\end{enumerate}
\end{thm:cor-os-buffer}

\begin{proof} We first run the algorithm from Theorem~\ref{thm:orth-sep} with $m = \nicefrac{2}{\delta}$ and obtain sets $X'$ and $Y'$. If set $X'$ satisfies the third condition: $\min_{\bar u \in X'} \mu(X'\setminus \Ball(\bar u, R))\leq \delta \mu(U)$, we return sets $(X,Y)=(X',Y')$. Otherwise, we return empty sets,
$(X,Y)=(\varnothing,\varnothing)$. By Theorem~\ref{thm:orth-sep}, 
$\Pr\{\bar u \in X\} \leq \alpha$ and $\Pr\{\bar u \in Y\} \leq \varepsilon \alpha$ for all $\bar u \in X$. Also, condition (3) always holds (because if $X'$ does not satisfy it, we return $\varnothing$). We now lower bound $\Pr\{\bar u\in X\}$:
\begin{align*}
\Pr\{\bar u\in X\} &= 
\Pr\{\bar u\in X'\} - 
\Pr\{\bar u\in X'\text{ and }
X=\varnothing \}\\
&= 
\Pr\{\bar u\in X'\} \cdot (1-
\Pr\{X = \varnothing \mid \bar u\in X'\}\\
&= \alpha(1 - \Pr\{X = \varnothing \mid \bar u\in X'\}).
\end{align*}
If $X = \varnothing$, then 
$$
\mu(X'\setminus \Ball(\bar u, R))\geq 
\min_{\bar v \in X'} \mu(X'\setminus \Ball(\bar v, R))> \delta \mu(U).$$
Thus,
$$\Pr\{X = \varnothing \mid \bar u\in X'\}\leq
\Pr\Big\{
 \mu(X'\setminus \Ball(\bar u, R)) >  \delta \mu(U)
\mid \bar u \in X'
\Big\}.
$$
However, by item (3) of Theorem~\ref{thm:orth-sep},
$$\E\Big[
 \mu(X'\setminus \Ball(\bar u, R)) 
\mid \bar u \in X'
\Big]\leq \frac{\mu(U)}{m} =
\frac{\delta\mu(U)}{2}.
$$
By Markov's inequality,
$$\Pr\{X = \varnothing \mid \bar u\in X'\}\leq \frac{1}{2}.$$
Therefore, $\Pr\{\bar u \in X \}\geq \alpha(1-1/2) = \alpha/2$. Finally, 
\begin{align*}
\Pr\{\bar v\notin X\cup Y \mid \bar u \in X\} &= 
\frac{\Pr\{\bar v\notin X\cup Y \text{ and } \bar u \in X\}}{\Pr\{\bar u \in X\}}\\
&=
\frac{\Pr\{\bar v\notin X'\cup Y' \text{ and } \bar u \in X'\}}{\Pr\{\bar u \in X'\}}
\cdot \frac{\Pr\{\bar u \in X'\}}{\Pr\{\bar u \in X\}}\\
&\leq 2\Pr\{\bar v\notin X'\cup Y' \mid \bar u \in X'\}\leq
2\calD\, \|\bar u - \bar v\|^2.
\qedhere
\end{align*}
\end{proof}

\newtheorem*{thm:os-two-buffers}{Theorem \ref{thm:orth-sep-two-buffers}}
\begin{thm:os-two-buffers}
    There exists a randomized procedure that given a finite set $U$ of unit vectors in $\bbR^d$ equipped with a measure $\mu$ and positive parameters $\varepsilon \in (0,1), \delta \leq 2/3, R \in (0,2)$, returns an $\delta$-orthogonal separator with two $\varepsilon$-buffers with distortion $\calD = O_R(\nicefrac{1}{\varepsilon}\;\log \nicefrac{1}{\delta})$, separation radius $R$, and probability scale $\alpha \geq O_R(1/poly(m))$. 

    For three disjoint random sets $X, Y, Z\subset U$ chosen from this orthogonal separator distribution, we have the following properties:
    \begin{enumerate}
        \item For all $\bar u \in U$, $\Pr\{\bar u \in X\} \in [\alpha/2, \alpha]$.
        \item For all $\bar u \in U$, $\Pr\{\bar u \in Y\} \leq \varepsilon \alpha$, and $\Pr\{\bar u \in Z\} \leq \varepsilon \alpha$.
        \item 
        $\min_{\bar u \in X} \mu(X\setminus \Ball(\bar u, R))\leq \delta \mu(U)$ (always).
        \item For all $\bar u, \bar v \in U$, 
        $\Pr\{\bar v\notin X\cup Y \mid \bar u \in X\} \leq \calD\, \|\bar u - \bar v\|^2$, and \\ $\Pr\{\bar v\notin X\cup Y\cup Z \mid \bar u \in X \cup Y\} \leq \calD\, \|\bar u - \bar v\|^2$,
        where $\calD = O_R(\nicefrac{1}{\varepsilon}\;\log \nicefrac{1}{\delta})$.
    \end{enumerate}
\end{thm:os-two-buffers}

\begin{proof}
We modify the algorithm in Theorem~\ref{thm:orth-sep} to generate three disjoint sets $X',Y',Z'$ as follows. We sample a $d$-dimensional Gaussian vector $g \sim \calN(0,I_d)$. For every vector $\bar{u}$ in $U$, we let $g_u = \langle\bar{u},g\rangle$ be the projection of vector $\bar{u}$ on the direction $g$. We use $\FFF(t)$ to denote the probability that a standard gaussian random variable is at least $t$.
We pick a threshold $t$ such that $\FFF(t) = \alpha$ for some $\alpha$ that we will specify later; our choice of $\alpha$ will guarantee that $t\leq 1$. Let $\varepsilon' = \varepsilon/ (e(t+1/t))$.  Then, we construct the orthogonal separator $X'$ and two buffers $Y', Z'$ as follows:
$$
X = \{\bar{u}: g_u \geq t\}; \qquad Y = \{\bar{u}: t - \varepsilon' < g_u < t \}; \qquad Z = \{\bar{u}: t-2\varepsilon' < g_u < t - \varepsilon'\}.
$$
If set $X'$ satisfies the third condition: $\min_{\bar u \in X'} \mu(X'\setminus \Ball(\bar u, R))\leq \delta \mu(U)$, we return sets $(X,Y,Z)=(X',Y',Z')$. Otherwise, we return empty sets,
$(X,Y,Z)=(\varnothing,\varnothing,\varnothing)$. 

By the similar analysis in Theorem~\ref{thm:orth-sep}, we have for all $\bar{u} \in U$, it holds that $\Pr\{\bar{u} \in X\} \leq \alpha$, $\Pr\{\bar{u} \in Y\} \leq \varepsilon\alpha$, and  $\Pr\{\bar{u} \in Z\} \leq \varepsilon\alpha$. By Theorem~\ref{thm:cor:orth-sep-main}, we have for all $\bar{u} \in U$, $\Pr\{\bar{u} \in X\} \geq \alpha/2$ and condition (3) always holds. Then, we show that condition (4) holds. The first part of condition (4) is the same as Theorem~\ref{thm:cor:orth-sep-main}. Note that $\alpha \leq \FFF(t-\varepsilon') \leq (1+\varepsilon)\alpha$. Using Lemma~\ref{lem:separate} on $g_u,g_v$ with parameters $\hat{m} = 1/\FFF(t-\varepsilon')$ and $\hat{\varepsilon} = \varepsilon'$, we have  
$$
\Pr\{g_u \geq t, g_v \leq t-\varepsilon' \} \leq O\left(\frac{\sqrt{\log \hat{m}}}{\varepsilon' \hat{m}}\right) \cdot\|\bar{u}-\bar{v}\|^2 \leq \alpha \calD \|\bar{u}-\bar{v}\|^2,
$$
where $\calD = O_R(\nicefrac{1}{\varepsilon}\log m)$.
\end{proof}

\bibliographystyle{alpha}
\bibliography{references}

\appendix

\section{Connection to Robust Expansion} \label{app:robustexp}

\newcommand{\epss}{\varepsilon^*}
In this section, we prove Corollary~\ref{corr:robustexp}.

\begin{proof}[Proof of Corollary~\ref{corr:robustexp}]
Let $\epss = \phi^V_\eta(G)$ be the robust vertex expansion of $G$. 
If $\epss = 0$, then the claim is trivial, because $\lambda_2 \geq 0$. So we assume below that $\epss > 0$.
Then for every disjoint subsets $S,T\subset V$ with $0<|S| \leq |V|/2$ and $|T| < \epss |S|$, we have 
\begin{equation}\label{eq:delta-S-T}
\delta(S, T) < (1-\eta) \delta(S,V\setminus S),
\end{equation}
as otherwise, we would have a contradiction 
$$\epss= \phi_\eta^V(G) \leq \phi_\eta^V(S) =\frac{N_\eta(S)}{|S|}\leq \frac{|T|}{|S|} < \epss.$$

Now we apply Corollary~\ref{cor:main-no-loss-in-k} of Theorem~\ref{thm:main} with $k=2$ and $\varepsilon' = \varepsilon^*/2$. We get an $\varepsilon'$-buffered partition $(P_1, P_2||B_1, B_2)$ with $\phi_G(P_1,P_2||B_1,B_2) \leq O(\lambda_2/\varepsilon')$.
Assume without loss of generality that $|P_1| \leq n/2$. Note that $|B_1| \leq \varepsilon'|P_1|  < \epss |P_1|$ and thus by~(\ref{eq:delta-S-T}), 
$$\delta(P_1, B_1) < (1-\eta) \delta(P_1,V\setminus P_1).$$
Therefore, 
$$\delta(P_1,V\setminus(P_1 \cup B_1)) = \delta(P_1,V\setminus P_1) - \delta(P_1,B_1) > \eta\, \delta(P_1,V\setminus P_1).$$
On the other hand, 
$$\delta(P_1,V\setminus(P_1 \cup B_1)) \leq d\cdot \phi_G(P_1,P_2||B_1,B_2) \cdot |P_1| \leq O\left(\frac{d\lambda_2 |P_1|}{\epss}\right).$$
We conclude that
$$\lambda_2 \geq \Omega(\eta) \cdot\epss\cdot \frac{\delta(P_1,V\setminus P_1)}{d|P_1|} = 
\Omega(\eta \cdot\phi^V_\eta(G)\cdot \phi_G(P_1)) \geq \Omega(\eta\cdot \phi^V_\eta(G)\cdot h_G).$$
\end{proof}

\section{Heavy Set 
\texorpdfstring{$P_t$}{P} in a Buffered Partition}\label{sec:heavy-set-Pt}
In this section, we argue why we may assume that one of the sets $P_t$ in the buffered partitioning $(P_1,\dots, P_k||B_1,\dots, B_k)$ contains at least $\Omega(\delta n)$ vertices (where $n=|V|$).
\begin{corollary}
There exists a buffered partitioning as in Theorem 1.1 (possibly with a different function $c(\delta)$ such that $|P_t| = \Omega(\delta n)$ for some $t$.
\end{corollary}
\begin{proof}
Let $\delta' = \sqrt{1+\delta} - 1 = \Theta(\delta)$ and $k' = \lfloor(1+\delta')k\rfloor$.
Apply Theorem~\ref{thm:main} with parameters $k'$ and $\delta'$. We get an $\varepsilon$-buffered partitioning $(P_1,\dots, P_{k'}\parallel B_1,\dots, B_{k'})$ with 
$$\phi_0 = \phi_G(P_1,\dots, P_{k'}\parallel B_1,\dots, B_{k'})
\leq \frac{c(\delta') \log k'}{\varepsilon} \lambda_{\lfloor(1+\delta)k\rfloor}.
$$
Assume without loss of generality that $|P_1|\leq |P_2| \leq \dots \leq |P_{k'}|$. Merge sets $P_k,\dots, P_{k'}$ and sets $B_k,\dots, B_{k'}$. That is, let $P'_k = \bigcup_{i=k}^{k'} P_i$ and $B'_k = \bigcup_{i=k}^{k'} B_i$. We obtain a buffered partitioning $(P_1,\dots, P_{k-1}, P'_k \parallel B_1, \dots, B_{k-1}, B_k')$. We show that it is $\varepsilon$-buffered and that its buffered expansion is at most $\phi_0$.
Clearly, merging does not change the value of $\phi_G(P_i \parallel B_i)$ for $i\in [k-1]$, as it does not change sets $P_i$ and $B_i$. So it is sufficient to verify that $|B_k'| \leq \varepsilon |P_k'|$ and $\phi_G(P'_k\parallel B'_k) \leq \phi_0$. Indeed,
\begin{align*}
|B_k'| &\leq \sum_{i=k}^{k'} |B_i| \leq \sum_{i=k}^{k'} \varepsilon|P_i|= \varepsilon |P_k'|.\\
\phi_G(P'_k\parallel B'_k) &= \frac{\delta_G(P_k', V\setminus (P_k'\cup B_k'))}{|P_k'|} \leq 
\frac{\sum_{i=k}^{k'}\delta_G(P_i, V\setminus (P_i\cup B_i))}{|P_k'|}
\leq 
\frac{\sum_{i=k}^{k'} \phi_0 |P_i|}{|P_k'|} = \phi_0.
\end{align*}
We used that sets $P_k,\dots, P_{k'}$ are disjoint and thus $|P_k'| = |P_k|+\dots + |P_{k'}|$.
Finally, we observe that $P_k'$ is the union of $k'-k+1 = \Omega(\delta k)$ largest sets out of $k'$ sets that together cover at least $(1-\varepsilon)n$ vertices. Thus, $|P_k'| \geq \frac{k'-k+1}{k'} (1-\varepsilon)n= \Omega(\delta n) $.
\end{proof}

\section{Lower Bound for \texorpdfstring{$k$}{k}-way Expansion and Pseudo-approximation Algorithm for Sparsest \texorpdfstring{$k$}{k}-way Partitioning}\label{sec:lb-and-approx}

In this section, we present the lower bound for non-buffered $k$-way expansion $h^k_G$ of graphs with vertex weights and edge costs. The proof is similar to that for graphs without vertex weights shown in~\cite{LouisRTV12,LeeOvT12}.
Combined with Theorem~\ref{thm:main-weighted}, it gives a pseudo-approximation alghorithm for the Sparsest $k$-way Partitioning problem.
\begin{proposition}\label{prop:lb}
Given any graph $G = (V, E, w, c)$ with vertex weights $w_u>0$ and edge costs $c_{uv}>0$, for any integer $k >1$, the $k$-way expansion is at least
$$
h^k_G \geq \frac{\lambda_k}{2}.
$$
\end{proposition}

\begin{proof}
    Let $P_1,P_2,\dots,P_k$ be the optimal solution for $k$-way expansion. Then, we have for any $i \in [k]$
    $$
    \phi_G (P_i) = \frac{|\delta(P_i,V\setminus P_i)|}{w(P_i)}\leq h^k_G.
    $$
    Let $\ONE_{P_i}$ be the indicator vector of set $P_i$ for all $i \in [k]$, i.e. $\ONE_{P_i}(u) = 1$ if $u\in P_i$, otherwise $\ONE_{P_i}(u) = 0$. Then, we use $x_{P_i} = D_w^{1/2}\ONE_{P_i}$ to denote the weighted indicator vector. Let $X= \{x_{P_i}: i\in [k]\}$. Since all vectors in $X$ are orthogonal to each other, the span of $X$ has dimension $k$. By the Courant-Fischer Theorem, we have
    \begin{equation}\label{eqn:lb_proof}
        \lambda_k = \min_{S \subset \bbR^n: \mathrm{dim}(S)=k} \max_{x \in S} \frac{x^T D_w^{-1/2}L_GD_w^{-1/2} x}{x^Tx} \leq \max_{x \in span(X)} \frac{x^T D_w^{-1/2}L_GD_w^{-1/2} x}{x^Tx}.
    \end{equation}
    Suppose $x \in span(X)$ is the maximizer of the right-hand side of Equation~(\ref{eqn:lb_proof}). We can write $x = \sum_{i=1}^k \alpha_i x_{S_i}$ for $\alpha_i \in \bbR$. Then, we have
    \begin{multline*}
        x^T D_w^{-1/2}L_GD_w^{-1/2} x = \left(\sum_{i=1}^k \alpha_i \ONE_{S_i}\right)^T L_G \left(\sum_{i=1}^k \alpha_i \ONE_{S_i}\right)=\\
        = \sum_{(u,v)\in E} c_{uv} \left(\sum_{i=1}^k \alpha_i \ONE_{S_i}(u)- \sum_{i=1}^k \alpha_i \ONE_{S_i}(v)\right)^2
        \leq 2\sum_{i=1}^k\alpha_i^2 \sum_{(u,v)\in E} c_{uv} (\ONE_{S_i}(u)-\ONE_{S_i}(v))^2,
    \end{multline*}
    where the last inequality is due to the relaxed triangle inequality, for any edge $(u,v)\in E$ with $u \in S_i$ and $v \in S_j$, $(\alpha_i\ONE_{S_i}(u) - \alpha_j\ONE_{S_j}(v))^2 \leq 2\alpha_i^2\ONE_{S_i}(u)^2 + 2 \alpha_j^2\ONE_{S_j}(v)^2$. Taking it into Equation~(\ref{eqn:lb_proof}), we have
    $$
    \lambda_k \leq \frac{2\sum_{i=1}^k\alpha_i^2 \sum_{(u,v)\in E} c_{uv} (\ONE_{S_i}(u)-\ONE_{S_i}(v))^2}{\sum_{i=1}^k \alpha_i^2 \sum_{u\in V} w_u \ONE_{S_i}(u)} = \frac{2\sum_{i=1}^k \alpha_i^2 |\delta(P_i,V\setminus P_i)|}{\sum_{i=1}^k \alpha_i^2 w(P_i)} \leq 2h^k_G.
    $$
\end{proof} 

Plugging the bound on $\lambda_{\lfloor (1+\delta)k\rfloor(L_G)}$ from Proposition~\ref{prop:lb} into Theorem~\ref{thm:main-weighted}, we get the following $O_{\varepsilon,\delta}(\log k)$ pseudo-approximation algorithm for the Sparsest $K$-Partitioning problem from 
\begin{theorem}
\label{thm:approx-sparsest-k-partitioning}
There exists a polynomial-time algorithm that given a graph $G = (V, E, w, c)$ with vertex weights $w_u>0$ and edge costs $c_{uv}>0$, $\varepsilon > 0$, $\delta > 0$, and $k >1$ such that $\max_{u\in V} w_u \leq \varepsilon w(V)/(3k)$, finds a $\varepsilon$-buffered partition $(P_1,\dots,P_k\parallel B_1,\dots, B_k)$ with 
$$
\phi_G(P_1,\dots,P_k\parallel B_1,\dots, B_k) \leq \frac{\kappa(\delta)\log k}{\varepsilon}h^{\lfloor(1+\delta)k\rfloor}_G.
$$
\end{theorem}
Note that in this theorem, we compare the cost of our $\varepsilon$-buffered $k$-partition to that of the optimal non-buffered $\lfloor (1+\delta)k\rfloor$-partition. 
\section{Buffered Balanced Cut}\label{sec:buf-balanced-cut}
In this section, we present our results for the buffered balanced cut. Consider any graph $G(V,E,w,c)$ with vertex weight $w_u>0$ and edge cost $c_{uv}>0$. For any $0 < \gamma \leq 1/2$, the $\gamma$-balanced cut of graph $G$ is a partition of graph $(L,R)$ such that $w(L), w(R) \in [\gamma w(V), (1-\gamma)w(V)]$. The $\gamma$-balanced cut problem asks to find a $\gamma$-balanced cut of a graph to minimize the cut size $\delta(L,R)$. We consider the $\varepsilon$-buffered $\gamma$-balanced cut. Given a weighted graph $G(V,E,w,c)$, the $\varepsilon$-buffered $\gamma$-balanced cut is a partition of graph $G$, $(L,R\parallel B)$ such that $w(L), w(R) \in [\gamma w(V), (1-\gamma)w(V)]$ and $w(B) \leq \varepsilon \min(w(L),w(R))$. We show a bi-criteria approximation for the balanced cut problem with an $\varepsilon$-buffered balanced cut. 

\begin{theorem}\label{thm:buffer-balanced}
Let $\varepsilon \in (0,1/4)$.
Consider any weighted graph $G=(V,E,w,c)$ with vertex weight $w_u > 0$ and $c_{uv} > 0$. There is a polynomial-time algorithm that finds three disjoint sets $L,B,R$ with $L \cup B \cup R = V$, $ w(L), w(R) \in [\nicefrac{1}{4} \cdot w(V),  \nicefrac{3}{4} \cdot w(V)]$, and $w(B) \le 3\varepsilon \min(w(L),w(R))$ such that  
$$
\delta(L,R) \leq O(1/\varepsilon) \cdot \delta(L^*,R^*),
$$
where $(L^*,R^*)$ is the optimal $1/3$-balanced cut.
$(L,R\parallel B)$ is a $(3\varepsilon)$-buffered $1/4$-balanced cut with cut size at most $O(1/\varepsilon)$ times the size of the optimal $1/3$-balanced cut.
\end{theorem}

\begin{proof}
We first describe our algorithm for buffered balanced cut, which is inspired by the approximation algorithm for balanced cut in~\cite{leighton1999multicommodity}. The algorithm recursively partitions the graph by using the buffered spectral partitioning algorithm in Section~\ref{sec:warm-up}. At the beginning, we set the graph $G_1 = G$. Then, we run the $\varepsilon$-buffered spectral partitioning to find a partition $(L_1,R_1 \parallel B_1)$ of the graph $G_1$. Suppose $w(L_1) \leq w(R_1)$. If $w(L_1) < w(V)/4$, then we recursively run the $\varepsilon$-buffered spectral partitioning on the subgraph $G_2$ of $G$ on the set of vertices $R_1$. For each call of buffered spectral partitioning, we label the partition $(L_t,R_t \parallel B_t)$ such that $w(L_t) \leq w(R_t)$. We recursively call the $\varepsilon$-buffered spectral partitioning until $\sum_{t=1}^T w(L_t) \geq w(V)/4$. Then, the algorithm returns the partition $(L,R,B)$ of $G$, where $L = \bigcup_{t=1}^T L_t$, $B = \bigcup_{t=1}^T B_t$, and $R = V \setminus (L \cup B)$. 

Then, we show that the partition $(L,R \parallel B)$ returned by this algorithm is a $3\varepsilon$-buffered $1/4$-balanced cut. Let $(L_t,R_t \parallel B_t)$ be the buffered partition of graph $G_t$ returned by the $t$-th call of the buffered spectral partitioning. Then, we have $w(L_t) \leq w(V_t)/2$ and $w(B_t) \leq \varepsilon w(L_t)$. 
Suppose the algorithm calls the buffered spectral partitioning for $T$ times. 
Then, we have $w(L) = \sum_{t=1}^{T} w(L_t) \geq w(V)/4$ and $\sum_{t=1}^{T-1} w(L_t) < w(V)/4$. Since $w(V_T) \leq w(V)$, we have 
$$
w(L) = \sum_{t=1}^{T} w(L_t) \leq \sum_{t=1}^{T-1} w(L_t) + w(L_T) \leq w(V)/4 + w(V_T)/2 \leq 3/4\cdot w(V).
$$ 
Since $w(L) \geq w(V)/4$, we have $w(R) \leq 3/4\cdot w(V)$. Since $w(L_T) \leq w(V_T) /2$ and $w(B_T)\leq \varepsilon w(L_T)$, we have 
$$
w(R) = w(V_T) - w(L_T)-w(B_T) \geq \left(1-\frac{1+\varepsilon}{2}\right) w(V_T).
$$
Note that $w(V_T) = w(V) - \sum_{t=1}^{T-1} w(L_t) + w(B_t) \geq \left(1-\frac{1+\varepsilon}{4}\right) w(V)$. Since $\varepsilon \leq 1/4$, we have
$$
w(R)\geq \left(1-\frac{1+\varepsilon}{2}\right) \left(1-\frac{1+\varepsilon}{4}\right) w(V) \geq \frac{w(V)}{4}.
$$
Thus, we have both $w(L)$ and $w(R)$ are in $[w(V)/4, 3w(V)/4]$. Since $w(B_t) \leq \varepsilon w(L_t)$ for all $t$, we have $w(B) \leq \varepsilon w(L)$ and
$$
w(B) \leq \varepsilon w(L) \leq \varepsilon\cdot 
\frac{3}{4} w(V) \leq 3\varepsilon \cdot w(R).
$$
Hence, we have $w(B) \leq 3\varepsilon \cdot \min \{w(L),w(R)\}$. 

Next, we bound the size of buffered cut $(L,B,R)$. For each call of the buffered spectral partitioning, we bound the cut size $\delta(L_t,R_t)$ for the buffered partition $(L_t,B_t,R_t)$ of graph $G_t$. 
Let $(L^*,R^*)$ be the optimal non-buffered $1/3$-balanced partition of graph $G$. Let $L^*_t = L^*\cap V_t$ and $R^*_t = R^* \cap V_t$. Then, we have $\delta(L^*_t,R^*_t) \leq \delta(L^*,R^*)$. Note that the weight of vertices in $V \setminus V_t$ is at most
$$
w(V\setminus V_t) = \sum_{i=1}^{t-1} w(L_i) + w(B_i) \leq (1+\varepsilon) \cdot \frac{w(V)}{4}.
$$
Suppose $w(L^*_t) \geq w(R^*_t)$. Since $w(L^*) \geq w(V)/3$ and $\varepsilon \leq 1/4$, we have
$$
w(L^*_t) \geq w(L^*) - w(V\setminus V_t) \geq \left(\frac{1}{3} - \frac{1+\varepsilon}{4}\right) w(V) \geq \frac{1}{48} w(V).
$$
By Proposition~\ref{prop:lb}, we have
$$
\frac{\lambda_2(L_{G_t})}{2} \leq \min_{S \subset V_t: w(S) \leq w(V_t)/2} \frac{\delta(S,V_t\setminus S)}{w(S)} \leq  \frac{\delta(L_t^*,R^*_t)}{w(L^*_t)}. 
$$
By Proposition~\ref{prop:cheeger}, we have 
\begin{multline*}
\delta(L_t,R_t) \leq 4\left(1+\frac{8}{\varepsilon}\right) \lambda_2(L_{G_t}) \cdot w(L_t) \leq \\
\leq 8\left(1+\frac{8}{\varepsilon}\right)\cdot \frac{w(L_t)}{w(L^*_t)} \cdot \delta(L^*_t,R^*_t) \leq 
O\left(\frac{1}{\varepsilon}\right) \cdot \frac{w(L_t)}{w(V)} \cdot \delta(L^*_t,R^*_t).
\end{multline*}
Combining all cuts edges in $\delta(L_t,R_t)$ for $T$ calls of buffered spectral partitioning, we have
$$
\delta(L,R) \leq \sum_{t=1}^T \delta(L_t,R_t) \leq O\left(\frac{1}{\varepsilon}\right) \cdot \sum_{t=1}^T \frac{w(L_t)}{w(V)} \cdot \delta(L^*_t,R^*_t) \leq O\left(\frac{1}{\varepsilon}\right) \delta(L^*_t,R^*_t),
$$
where the last inequality is due to $w(L) \leq \nicefrac{3}{4}\cdot w(V)$.
\end{proof}


We also consider the $k$-way balanced partition problem. Given a graph $G(V,E,w,c)$, for any $\gamma \geq 1$, we say that $P_1,P_2,\dots,P_k$ is a $(\gamma,k)$-balanced partition 
 of $G$  if $w(P_i) \leq \gamma w(V)/k$ for all $i\in [k]$. The $(\gamma,k)$-balanced partition problem aims to find a $(\gamma, k)$ balanced partition to minimize the total cost of edges with two endpoints in different parts. By using the buffered balanced cut algorithm in Theorem~\ref{thm:buffer-balanced} and the recursive bi-section algorithm in~\cite{simon1997good}, we show a bi-criteria approximation for the $k$-way balanced partition.

\begin{corollary}\label{cor:approx-k-balanced}
Let $\varepsilon \in (0,1/4)$.
Consider any weighted graph $G=(V,E,w,c)$ with vertex weight $w_u > 0$ and $c_{uv} > 0$. There is a polynomial-time algorithm that finds a $\varepsilon$-buffered $(6,k)$-balanced partition $P_1,P_2,\dots,P_k, B$ such that $P_1,P_2,\dots,P_k$ and $B$ are disjoint, $w(B) \leq O(\varepsilon) w(V)$, and
$$
\sum_{i<j} \delta(P_i, P_j) \leq O(\nicefrac{1}{\varepsilon}\cdot\log^2 k)\cdot \mathrm{OPT},
$$
where $\mathrm{OPT}$ is the optimal cost for $(1,k)$-balanced partition. 
\end{corollary}
\section{Graphs with Vertex Weights and Edge Costs}\label{sec:reduction}
In this section, we prove our main results for graphs $G=(V,E, w, c)$ with vertex weights $w_u > 0$ and edge costs $c_{uv} > 0$.

Theorem~\ref{thm:main} holds for regular graphs with parallel edges but without edge costs and vertex weights. Assume that we have a graph $G$ with edge costs $c_{uv}$ and with vertex weights $w_u = 1$ such that the total cost of all edges incident on a vertex does not depend on the vertex; that is, $C_0 = \sum_{v:(u,v)\in E} c_{uv}$ does not depend on $u$. If all edge costs are integers, we can simulate edge costs by adding parallel edges -- we replace each edge $(u,v)$ with $c_{uv}$ parallel edges. We obtain a $C_0$-regular graph $G'$. Let $L_{G'} = I-\frac{1}{C_0}A_{G'}$ be the normalized Laplacian of $G'$. Let $\Lap_G = D_w^{-1/2} \nL_G D_w^{-1/2}$ be the normalized Laplacian of $G$. It is immediate that $\Lap_G = C_0 \Lap_{G'}$ and $\delta_{G}(A,B) = \delta_{G'}(A,B)$ for every $A, B\subseteq V$. 
Let $k' = \lfloor (1+\delta) k\rfloor$. Then, $\lambda_{k'}(\Lap_{G'}) = \lambda_{k'}(\Lap_G)/C_0$. 

By Theorem~\ref{thm:main}, there exists an $\varepsilon$-buffered partition $(P_1,\dots, P_k\parallel B_1, \dots, B_k)$ such that 
$$
\phi_{G'}(P_i \parallel B_i) = \frac{\delta_{G'}(P_i, V\setminus (P_i \cup B_i))}{C_0 |P_i|} \leq \frac{c(\delta) \log k}{\varepsilon} \cdot \lambda_{k'}(\Lap_{G'})
$$
for every $i\in [k]$.
Since $\lambda_{k'}(\Lap_{G'}) = \lambda_{k'}(\Lap_G)/C_0$ and $w(P_i) = |P_i|$, we have for all $i$,
\begin{equation}\label{eq:reduction-first-step}
\phi_{G}(P_i \parallel B_i) = \frac{\delta_{G}(P_i, V\setminus (P_i \cup B_i))}{w(P_i)} \leq \frac{c(\delta) \log k}{\varepsilon} \cdot \lambda_{k'}(\Lap_{G}).
\end{equation}
Now if we multiply all edge costs by the same positive number $\rho$, both the left and right hand side will get multiplied by $\rho$. Therefore, the inequality holds not only for integer edge costs but also for arbitrary positive rational costs. By continuity, it holds for arbitrary positive edge costs. We get the following corollary.
\begin{corollary}
Let $G$ be a graph with positive edge costs $c_{uv}$ and unit vertex weights such that $C_0 = \sum_{v:(u,v)\in E} c_{uv}$ is the same for all vertices $u$. Then there exists an $\varepsilon$-balanced partition $(P_1, \dots, P_k \parallel B_1,\dots, B_k)$ such that inequality (\ref{eq:reduction-first-step}) holds for all $i$.
\end{corollary}

Now we present a black-box reduction that  proves Theorem~\ref{thm:main-weighted}.
We note that the reduction can significantly increase the running time of the algorithm. However, in fact, we can use the algorithm from Theorem~\ref{thm:main} to find $(P_1,\dots, P_k\parallel B_1,\dots, B_k)$ (the proof of this fact essentially repeats that of Theorem~\ref{thm:main}, and we do not present it here). 
\begin{theorem}\label{thm:main-weighted-proof}
Let $G = (V, E, w, c)$ be a graph with positive weights $w_u > 0$ and edge costs $c_{uv} > 0$, $\varepsilon \in [0,1)$, $\delta \in (0,1)$, and $k\geq 2$ be an integer. Assume that $\max_u w_u \leq \varepsilon w(V) /(3k)$.
Let $\Lap_G = D_w^{-1/2} \nL_G D_w^{-1/2}$ be the normalized Laplacian of $G$. Then
\begin{equation}
\label{eq:Cheeger-weighted}
  h_G^{k,\varepsilon} \leq \frac{\kappa(\delta) \log k}{\varepsilon} 
   \cdot \lambda_{\lfloor(1+\delta)k\rfloor}(\Lap_G),
\end{equation}
where $\kappa(\delta)$ is a function that depends only on $\delta$.
%
\end{theorem}
\begin{proof}
Assume first that all vertex weights  are integers greater than or equal to $2$. Let $W = \sum_{u \in V} w_u$ be the total weight of all vertices. Let $C = \sum_{(u,v)\in E} c_{uv}$ be the total cost of all edges and $B = C\cdot W^2$.

We construct an auxiliary graph $G'$ with unit vertex weights as follows.  
For each vertex $u$ of $G$, we create its own ``cloud of vertices'' $Q_u$ of size $w_u$; all vertices $q\in Q_u$ have unit weights. 
For $(u,v)\in E$, we connect every $q\in Q_u$ with every $q'\in Q_v$ by an edge $(q,q')$ with cost $c'_{qq'} = \frac{c_{uv}}{|Q_u||Q_v|}$. Note that the total cost of all edges between $Q_u$ and $Q_v$ equals $c_{uv}$. Let $b_u =  \sum_{v:(u,v)\in E} \frac{c_{uv}}{|Q_u|}$  be the total cost of edges incident on vertex $q\in Q_u$ (so far). Now we connect every two vertices $q,q'\in Q_u$ by an edge of cost $c'_{qq'} = \frac{B - b_u}{|Q_u| - 1}$. After this step, the total cost of all edges incident on $q\in Q_u$ is exactly $B$, since $q$ has $|Q_u| - 1$ neighbors in $Q_u$. We denote the obtained graph by $G'$.

\paragraph{Properties of $G'= (V', E')$ that we established.}
\begin{itemize}
    \item $|Q_u| = w_u$; all vertices have unit weights in $G'$.
    \item The total cost of all the edges between $Q_u$ and $Q_v$ is $c_{uv}$.
    \item The total cost of all edges incident on every vertex equals $B$ (and does not depend on $u$).
    \item $G'[Q_u]$ is a clique, in which all edges have cost $c_{qq'}' =\frac{B - b_u}{|Q_u| - 1} \geq \frac{B-C}{W-1} > CW$.
\end{itemize}

\noindent Now we upper bound $\lambda_{k'}(\Lap_{G'})$ in terms of $\lambda_{k'}(\Lap_G)$.
\begin{lemma} $\lambda_{k'}(\Lap_{G'}) \leq \lambda_{k'}(\Lap_G)$
\end{lemma}
\begin{proof}
Let $x_1,\dots,x_{k'}$ be the first $k'$ orthogonal unit eigenvectors of $\Lap_G$.
Define vectors $z_1,\dots, z_{k'} \in {\mathbb R}^{|V'|}$ as follows: for $q\in Q_u$, we let $z_i(q) = \frac{x_i(u)}{\sqrt{w_u}}$.
First, observe that $z_1,\dots, z_{k'}$ are pairwise orthogonal unit vectors:
\[
\langle z_i, z_j\rangle = \sum_{q\in V'} z_i(q)z_j(q) = \sum_{u\in V}\sum_{q\in Q_u} z_i(q)z_j(q) = \sum_{u\in V} |Q_u| \frac{x_i(u)x_j(u)}{w_u} = \langle x_i, x_j\rangle = 
\begin{cases}
0, & \text{ if } i\neq j\\
1, & \text{ if } i = j
\end{cases}
\]
Further,
\begin{align*}
z_i^T \Lap_{G'} z_j&= \sum_{(q,q')\in E'} c'_{qq'}(z_i(q) - z_i(q'))\cdot (z_j(q) - z_j(q')) \\
&= 
\sum_{(u,v)\in E}\sum_{\substack{q\in Q_u\\q'\in Q_v\\(q,q')\in E'}} \frac{c_{uv}}{|Q_u||Q_v|} (z_i(q) - z_i(q'))\cdot (z_j(q) - z_j(q')) \\
&\qquad + \sum_{u \in V} \frac{B-b_u}{|Q_u| - 1}\sum_{\substack{q,q'\in Q_u\\(q,q')\in E'}} (z_i(q) - z_i(q'))\cdot (z_j(q) - z_j(q'))\\
&
=\sum_{(u,v)\in E} c_{uv}
\left(\frac{x_i(u)}{w_u^{1/2}} - \frac{x_i(v)}{w_v^{1/2}}\right)\cdot \left(\frac{x_j(u)}{w_u^{1/2}} - \frac{x_j(v)}{w_v^{1/2}}\right)
= x_i^T \Lap_G x_j.
\end{align*}
We conclude that $z_i^T \Lap_{G'} z_j =\lambda_i(\Lap_G)$ if $i=j$, and $z_i^T \Lap_{G'} z_j = 0$, otherwise.

Finally, we use the Courant--Fischer theorem to upper bound $\lambda_{k'}(\Lap_G)$.
Let $H$ be the linear span of vectors $z_1,\dots, z_{k'}$. 
By the Courant--Fischer theorem,
\begin{align*}
\lambda_{k'}(\Lap_{G'}) &\leq \max_{z\in H\setminus\{0\}}\frac{z^T \Lap_{G'}z}{\|z\|^2}=
\max_{\substack{z=\sum_i \alpha_i z_i\\ \alpha\in {\mathbb R}^{k'}\setminus\{0\}}}
\frac{z^T \Lap_{G'}z}{\|z\|^2} = 
\max_{\alpha\in {\mathbb R}^{k'}\setminus\{0\}}
\frac{\sum_{i,j}(\alpha_i\alpha_j) z_i^T \Lap_{G'}z_j}{\|\alpha\|^2} \\&= 
\max_{\alpha\in {\mathbb R}^{k'}\setminus\{0\}}
\frac{\sum_{i}\alpha_i^2 \lambda_i(\Lap_G)}{\|\alpha\|^2} = 
 \lambda_{k'}(\Lap_G).
\end{align*}
\end{proof}

Let $\varepsilon' = \varepsilon/10$. We apply Theorem~\ref{thm:main} to ${G'}$ and obtain an $\varepsilon'$-buffered partition $(P_1',\dots, P_k' \parallel B_1',\dots, B_k')$ of $G'$ with 
$\phi_{G'}(P_1', \dots, P_k' \parallel B_1', \dots, B_k') \le \frac{c(\delta) \log k}{\varepsilon'}  \lambda_{k'}(\Lap_{G'}) \leq \frac{c(\delta) \log k}{\varepsilon'}  \lambda_{k'}(\Lap_G)$.
Observe that if some set $Q_u$ contains a vertex $q\in P_i'$ and a vertex $q'\in P_j' \cup B_j'$ with $j\neq i$ then $\phi_{G'}(P_i' \parallel B_i')$ is very large 
$$
\phi_{G'}(P_i' \parallel B_i') \geq \frac{\delta_{G'}(P_i', P_j' \cup B_j')}{w(P_i')} \geq \frac{c_{qq'}}{W} > C.
$$
Then, any partition $(P_1, \dots, P_k\parallel \varnothing,\dots, \varnothing)$ of $G$ satisfies the condition of the theorem:
$$\phi_G(P_i\parallel \varnothing) \leq\nicefrac{C}{2} < \phi_{G'}(P_i' \parallel B_i') \leq \frac{c(\delta) \log k}{\varepsilon'}  \lambda_{k'}(\Lap_G),$$
as required. 
So we assume below that if $P_i'\cap Q_u \neq \varnothing$ then $(P_j' \cup B_j') \cap Q_u = \varnothing$ for every $u$, $i$, and $j\neq i$. Then for every $u$, there are two possibilities: either
\begin{enumerate}
    \item $Q_u \subseteq P_i' \cup B_i'$ for some $i$, or
    \item $Q_u \subseteq \bigcup_i B_i'$.
\end{enumerate}
Depending on which of the possibilities takes place, we say that $u$ is a vertex of the first or second type, respectively\footnote{If $Q_u \subseteq B_i'$, let us assume that $u$ is of the first type.}.
Now we define an $\varepsilon$-buffered partition $(P_1,\dots, P_k \parallel B_1,\dots, B_k)$ of $G$.
First, we assign every vertex $u$ to one of the sets $P_1, \dots, P_k, B_1, \dots, B_k$ and $U$, where $U$ is a special set that will be partitioned among $B_1,\dots B_k$ later. We do that as follows: 
\begin{enumerate}
    \item if $|Q_u \cap P_i'| \geq |Q_u|/2$, we assign $u$ to $P_i$;
    \item otherwise, if $|Q_u \cap B_i'| \geq |Q_u| /2$, we assign $u$ to $B_i$;
    \item otherwise, we assign $u$ to $U$.
\end{enumerate}
Note that each vertex of the first type is necessarily assigned to some $P_i$ or $B_i$. Each vertex of the second type is assigned to some $B_i$ or $U$. 

Since $U$ consists of the vertices of the second type, we have $\bigcup_{u\in U}Q_u \subset \bigcup_i B_i'$ and thus
$$w(U) = \Bigl|\bigcup_{u\in U}Q_u\Bigr| \leq \Bigl|\bigcup_i B_i'\Bigr| \leq \varepsilon' \Bigl|\bigcup_i P_i'\Bigr|.$$
Here we used that that partition $(P_1', \dots, P_k' \parallel B_1', \dots, B_k')$ is $\varepsilon'$-buffered.
We create sets $B_1'', \dots, B_k''$, which are initially empty, and set the capacity of $B_i''$ to $\frac{\varepsilon |P_i'|}{2}$. 
We distribute vertices from $U$ one-by-one among $B_1'', \dots, B_k''$ so that the total weight assigned to $B_i'$ does not exceed its capacity. We stop when we either assign all the vertices from $U$ or no unassigned vertex in $U$ can be assigned to any $B_i''$, without violating the capacity requirement for $B_i''$.
We now show that this procedure assigns all the vertices from $U$. Indeed,
assume that some vertex $u$ is not assigned. Then, $w_u$ is greater than the the remaining capacity of every $B_i''$; that is, $w_u > \frac{\varepsilon |P_i'|}{2}  - w(B_i'')$ for every $i$. Adding up these inequalities over all $i$, 
we get 
\begin{align*}
    k w_u &> \sum_{i=1}^k\left(\frac{\varepsilon |P_i'|}{2} - w(B_i'')\right)
\geq \frac{\varepsilon}{2}\Bigl|\bigcup_i P_i'\Bigr| - w(U)
\geq \frac{\varepsilon}{2}\Bigl|\bigcup_i P_i'\Bigr| - \Bigl|\bigcup_i B_i'\Bigr|\\
&\geq \frac{\varepsilon}{2}\Bigl|\bigcup_i P_i'\Bigr| - \varepsilon' \Bigl|\bigcup_i P_i'\Bigr| \geq \frac{2\varepsilon}{5} \Bigl|\bigcup_i P_i'\Bigr|
\stackrel{\star}{\geq} \frac{2\varepsilon(1-\varepsilon')}{5} w(V) \geq \frac{\varepsilon w(V)}{3}
\end{align*}
Inequality {\tiny $\stackrel{\star}{\geq}$} above follows from two inequalities: $|\bigcup_i P_i'| + |\bigcup_i B_i'| = w(V)$ and $|\bigcup_i B_i'|\leq \varepsilon' |\bigcup_i P_i'|$. We get that $w_u > \frac{\varepsilon w(V)}{3k}$, which contradicts to the assumption of the theorem.
We conclude that $\bigcup_i B_i'' = U$. Finally, we add vertices from $B_i''$ to $B_i$ for every $i$. We obtain the desired partition $(P_1,\dots, P_k \parallel B_1,\dots, B_k)$.

Now we prove that $(P_1,\dots, P_k\parallel B_1,\dots, B_k)$ satisfies the desired requirements.
Fix $i$. We upper bound $\delta_G(P_i, V\setminus(P_i \cup B_i))$. Note that if edge $(u,v)$ goes from $P_i$ to $V\setminus(P_i \cup B_i)$ then $u$ is a vertex of the first type and $|Q_u\cap P_i'| \geq |Q_u|/2$ and either 
\begin{itemize} 
\item $v$ is a vertex of the first type and $Q_v \subseteq P_j \cup B_j$ for some $j\neq i$, or
\item $v$ is a vertex of the second type and at least one half of the vertices in $Q_v$ are not in $B_i$ (and none of them are in $P_i$). 
\end{itemize}
To summarize, in either case at least a half of the vertices in $Q_u$ lie in $P_i'$ and at least half of vertices in $Q_v$ do not lie in $P_i' \cup B_i'$. Thus, at least one quarter of all edges from $Q_u$ to $Q_v$ contribute to $\delta_{G'}(P_i', V'\setminus (P_i', B_i'))$, and their total contribution is at least $c_{uv}/4$.
We conclude that 
$$\delta_G(P_i, V\setminus(P_i \cup B_i)) \leq 4 \delta_{G'}(P_i', V\setminus(P_i' \cup B_i')).$$
Now we lower bound $w(P_i)$. Let $A$ be the set of vertices $u$ of the first type such that $Q_u \subseteq P_i'\cup B_i'$. Note that $P_i \subseteq A$
and $P_i' \subseteq \bigcup_{u\in A} Q_u$.
Consider $u\in A$. If $u\in P_i$, then $w(P_i \cap \{u\}) = |Q_u| \geq |Q_u\cap P_i'| - |Q_u \cap B_i'|$. If $u\notin P_i$, then $w(P_i \cap \{u\}) = 0 \geq |Q_u\cap P_i'| - |Q_u \cap B_i'|$, since $|Q_u\cap P_i'| < |Q_u|/2 \leq |Q_u \cap B_i'|$. We have,
$$w(P_i) = \sum_{u\in A} w(P_i \cap \{u\}) \geq \sum_{u\in A} |Q_u\cap P_i'| - |Q_u \cap B_i'| \geq |P'_i| - |B_i'| \geq (1 - \varepsilon') |P_i'|.$$
We have, 
$$\phi_G(P_i\parallel B_i) = \frac{\delta_G(P_i, V\setminus(P_i \cup B_i)}{w(P_i)} \leq \frac{4}{1-\varepsilon'}
\frac{\delta_{G'}(P_i', V\setminus(P_i' \cup B_i')}{|P_i'|} =  \frac{O(c(\delta)) \log k}{\varepsilon}  \lambda_{k'}(\Lap_G).
$$
It remains to show that partition $(P_1,\dots, P_k \parallel B_1, \dots, B_k)$ is $\varepsilon$-buffered.
We already showed that $w(P_i) \geq (1 - \varepsilon') w(P_i')$. Now we upper bound $w(B_i)$. 
First, $w(B_i'') \leq \nicefrac{\varepsilon|P_i'|}{2} \leq \nicefrac{\varepsilon w(P_i)}{2(1- \varepsilon')} \leq \nicefrac{5\varepsilon w(P_i)}{9}$.

Then, $u\in B_i\setminus B_i''$ if and only if $|Q_u \cap B_i'| \geq |Q_u|/2 = w_u/2$. Therefore, 
$$w(B_i \setminus B_i'') \leq 2\sum_{u\in B_i \setminus B_i''} |Q_u \cap B_i'| \leq 2|B_i'| \leq 2\varepsilon' |P_i'| \leq \frac{2\varepsilon'}{1 - \varepsilon'} w(P_i) \leq \frac{2\varepsilon}{9} w(P_i).$$
We conclude that $w(B_i) = w(B_i\setminus B_i'') + w(B_i'') = \frac{7\varepsilon}{9} w(P_i)$, as required. This completes the proof for the case when all the vertex weights are integers. By linearity, inequality (\ref{eq:Cheeger-weighted}) also holds when all the weights are rational numbers, and by continuity, it follows that inequality (\ref{eq:Cheeger-weighted}) holds when weights are arbitrary positive real numbers.
\end{proof}
\section{Lower Bound on \texorpdfstring{$h_G^{k,\varepsilon}$}{h(k,epsilon)}} \label{sec:lowerbound}
In this section, we prove Theorem~\ref{thm:lower-bound-hkG}, which we now restate as follows.
\begin{theorem}
Consider a $d$-regular graph $G=(V,E)$ and its $\varepsilon$-buffered partition $(P_1, \dots, P_k||B_1,\dots, B_k)$.
Then for every $i\in [k]$,
$$\lambda_k \leq 2\phi_G(P_1,\dots,P_k||B_1,\dots,B_k) + \varepsilon.$$
Thus,
$$\lambda_k \leq 2h^{k,\varepsilon}_G + \varepsilon.$$
\end{theorem}
\begin{proof}
By the Courant-Fischer min-max theorem,
$$\lambda_k = \min_H \max_{z \in H:z\neq 0} \frac{z^TL_Gz}{\|z\|^2},$$
where the minimum is over $k$-dimensional subspaces $H$ of ${\mathbb R}^n$. Let
$b_i$ be the indicator vector of $P_i$: $b_{i}(u) = 1$ if $u\in P_i$ and $b_{i}(u) = 0$, otherwise. Let $H$ be the linear span of $b_1,\dots,b_k$ and $z=\sum_{i=1}^k\alpha_i b_i$. Then,
$$\lambda_k \leq \max_{(\alpha_1,\dots,\alpha_k)\neq 0} \frac{z^TL_Gz}{\|z\|^2}$$
First note that vectors $b_i$ have disjoint supports and thus are mutually orthogonal. Therefore, $\|z\|^2 = \sum_{i=1}^k \alpha_i^2 \|h_i\|^2 = 
\sum_{i=1}^k \alpha_i^2 |P_i|$.
Now we upper bound $z^TL_G z$. We will use that 
$|\delta(P_i, B_i)| \leq d |B_i| \leq \varepsilon d |P_i|$.
\begin{align*}
dz^TL_Gz &
\stackrel{\tiny\text{by~(\ref{eq:lapl})}}{=}
\sum_{\substack{i,j\in[k]\\i< j}} (\alpha_i - \alpha_j)^2 \cdot |\delta(P_i,P_j)| + \sum_{i\in [k]}(\alpha_i - 0)^2 \cdot \bigl|\delta\bigl(P_i,\bigcup_j B_j\bigr)\bigr| \\
&\leq \sum_{\substack{i,j\in[k]\\i< j}} (2\alpha_i^2 + 2\alpha_j^2) \cdot |\delta(P_i,P_j)| + \sum_{i\in [k]}\left(\alpha_i^2 \cdot \bigl|\delta\bigl(P_i,\bigcup_{j:j\neq i}B_j \setminus B_i\bigr)\bigr| + \alpha_i^2 \cdot |\delta(P_i,B_i)|\right) \\
&=
2\sum_{\substack{i,j\in[k]\\i\neq j}} \alpha_i^2 \cdot |\delta(P_i,P_j)| + \sum_{i\in [k]}\alpha_i^2 \cdot\left( \bigl|\delta\bigl(P_i,\bigcup_{j:j\neq i}B_j \setminus B_i\bigr)\bigr| +  |\delta(P_i,B_i)|\right)\\
&\leq 
\sum_{i\in[k]} \alpha_i^2\cdot\left(2\left|\delta\Bigl(P_i, \bigcup_{j:j\neq i} P_j \cup 
\Bigl(\bigcup_{j:j\neq i}B_j \setminus B_i\Bigr)\Bigr)\right| + |\delta(P_i,B_i)|\right)\\
&\leq \sum_{i\in[k]}\alpha_i^2 \cdot\Bigl(2\bigl|\delta(P_i, V\setminus (P_i \cup B_i))\bigr| +  \varepsilon d |P_i|\Bigr).
\end{align*}
Therefore,
$$\frac{z^TL_Gz}{\|z\|^2} \leq 
\frac{1}{d}\max_{i\in[k]}
\frac{2\bigl|\delta(P_i, V\setminus (P_i \cup B_i))\bigr| +  \varepsilon d |P_i|}{|P_i|} = 
\max_{i\in[k]}
\frac{2\bigl|\delta(P_i, V\setminus (P_i \cup B_i))\bigr|}{d|P_i|} + \varepsilon.
$$
\end{proof}
\section{Gaussian Distribution}\label{sec:gaussian}
In this section, we present several useful estimates on the Gaussian distribution. Let $X\sim \calN(0,1)$ be a one-dimensional Gaussian
random variable. Denote the probability that $X\geq t$ by $\bar\Phi(t)$:
$$\bar\Phi(t) = \Pr\{X\geq t\}.$$
The first lemma gives an accurate estimate on $\bar\Phi(t)$ for large $t$.
\begin{lemma}(see \cite[Lemma A.1]{CMM2})\label{lem:NormalDistrib}
For every $t > 0$,
$$  \frac{t}{\sqrt{2\pi}\,(t^2+1)} e^{-\frac{t^2}{2}} < \bar{\Phi}(t) <
\frac{1}{\sqrt{2\pi}\,t}e^{-\frac{t^2}{2}} \text{ and } \FFF(t) = \Theta\Bigl(\frac{e^{-\frac{t^2}{2}}}{t+1}\Bigr).$$
\end{lemma}

\begin{lemma}\label{lem:Gaussian_1}
(see \cite[Lemma A.1, part 2]{CMM2})
    For any $\rho \geq 1$ and $t \geq 0$, there exists a constant $C$ such that 
    $$
    \FFF(\rho t) \leq \frac{1}{t} (Ct\FFF(t))^{\rho^2}.
    $$
\end{lemma}

\begin{lemma}\label{lem:separate}
Let $X$ and $Y$ be jointly $\calN(0,1)$-Gaussian random variables. Denote $\delta^2 = \nicefrac{1}{2}\Var[X - Y]$. Choose $m > 3$, threshold $t>1$ such that $\FFF(t) = 1/m$, and a parameter $\varepsilon\in [0, t]$.
Then
$$\Pr\{X \geq t \text{ and } Y \leq t - \varepsilon\} \leq O(\delta^2 \varepsilon^{-1} \sqrt{\log m}/m).$$
\end{lemma}
\begin{proof} Note that (1) the covariance of $X$ and $Y$ is $\E[XY] = 1-\Var[X-Y]/2 = 1-\delta^2$, and (2)~by Lemma~\ref{lem:NormalDistrib}, $t = \Theta(\sqrt{\log m})$.
Denote $p = \Pr\{X \geq t \text{ and } Y \leq t - \varepsilon\}$.
Note that if $\delta^2 \varepsilon^{-1} t \geq 1/32$, then the lemma trivially holds, 
$$p = \Pr\{X \geq t \text{ and } Y \leq t - \varepsilon\} \leq 
\Pr\{X \geq t\} = \frac{1}{m} \leq 
O\Bigl(\frac{\delta^2 \varepsilon^{-1} \sqrt{\log m}}{m}\Bigr),$$
as required. 
So we assume below that $\varepsilon > 32\delta^2 t$.
Let $\alpha = \E[XY] = 1-\delta^2$. Consider Gaussian random variable $Z = \alpha X - Y$. Note that $Z$ has mean 0 and variance $\E[Z^2] = \alpha^2 + 1 - 2\alpha^2 = 2\delta^2-\delta^4$. Further, the covariance of $X$ and $Z$ is 0: $\E[XZ] = \alpha\E[X^2] - E[XY]= 0$. 
In particular, for every $\tau \geq 0$, 
\begin{equation}\label{ineq:Z-tail}
\Pr\{Z \geq \tau\} = \FFF(\tau / \sqrt{2\delta^2 - \delta^4}) \leq \FFF\left(\frac{\tau}{\sqrt{2}\delta}\right)
\stackrel{\text{\tiny by Lemma~\ref{lem:NormalDistrib}}}\leq O\left(e^{-(\frac{\tau}{\sqrt{2}\delta})^2/2}\right).
\end{equation}
Therefore, $X$ and $Z$ are independent. We have, 
$$
p = \Pr\{X \geq t \text{ and } Y \leq t - \varepsilon\} = 
\Pr\{X \geq t \text{ and } \alpha X - Z \leq t - \varepsilon\} =
\frac{1}{m}\Pr\{Z \geq \varepsilon + \alpha X - t\given X \geq t\} 
$$
Define random variable $\Delta = X - t$. Then 
$$\varepsilon + \alpha X - t=
\varepsilon + (1 - \delta^2) (t + \Delta) - t \geq 
\varepsilon/2 + (1 - \delta^2) \Delta \geq \frac{\varepsilon + \Delta}{2},$$
where we used that $\varepsilon/2 - \delta^2 t \geq 0$ and $\delta^2\leq \varepsilon/(2t) \leq t/(2t) = 1/2$. We have,
$$
p\leq  
\frac{1}{m}\Pr\{Z \geq (\varepsilon  + \Delta)/2 \given \Delta \geq 0\}
\stackrel{\text{\tiny by (\ref{ineq:Z-tail})}}\leq
\frac{\E[e^{-(\frac{\varepsilon+\Delta}{2})^2/(4\delta^2)} \given \Delta \geq 0]}{m}.
$$
Let us upper bound the probability density function $f_\Delta(x)$ of $\Delta$ conditioned on the event $\Delta \geq 0$. 
\begin{align*}f_\Delta(x) &=  
\left. \frac{e^{-(x+t)^2/2}}{\sqrt{2\pi}}\right/\Pr\{\Delta \geq 0\} = 
(t+1)\cdot \frac{e^{-t^2/2}}{\sqrt{2\pi} (t+1)}\cdot e^{-x^2/2-tx}\cdot m\\
&\leq 
O\bigl(t \cdot \FFF(t) \cdot e^{-x^2/2-tx}\cdot m\bigr) = O\bigl(t  e^{-x^2/2-tx}\bigr) = O\bigl(t  e^{-tx}\bigr).
\end{align*} 
We conclude that
\begin{align*}
pm &\leq O(1) \int_{0}^\infty   e^{\frac{-(\varepsilon  + x)^2}{16\delta^2}} (te^{-tx}) dx
=
O\Bigl(te^{4t^2\delta^2+\varepsilon t}\Bigr) \int_{0}^\infty  e^{\frac{-(x  + 8t \delta^2 + \varepsilon)^2}{16\delta^2}} dx\\
&\stackrel{\tiny \text{let } \tilde x = \frac{x+8t\delta^2 + \varepsilon}{2\sqrt{2}\delta}} =
O\Bigl(\delta t e^{4t^2\delta^2+\varepsilon t}\Bigr) \int_{8t\delta^2 + \varepsilon/(2\sqrt{2}\delta)}^\infty  e^{\frac{-\tilde x^2}{2}}d \tilde x
\leq O\Bigl(\delta t e^{2\varepsilon t} \FFF\Bigl(\frac{\varepsilon}{2\sqrt{2}\delta}\Bigr)\Bigr)
\\
&\stackrel{\text{\tiny by Lemma~\ref{lem:NormalDistrib}}}{\leq} 
O\left(\frac{\delta^2}{\varepsilon} t e^{2\varepsilon t - \varepsilon^2/(16\delta^2)}
\right) = O\left(\frac{\delta^2t}{\varepsilon}\right)
\end{align*}
here we three times used that $\varepsilon > 32\delta^2 t$.
We conclude that $p = O(\delta^2 \varepsilon^{-1} t/m) =
O(\delta^2 \varepsilon^{-1} \sqrt{\log m}/m)$, as required.
\end{proof}

\newpage

\section{Supplementary Figures}

\begin{figure}[h!]%
    \centering
    \subfloat[\centering Buffered partition $S, B, T = V \setminus (S \cup B)$.]{{\includegraphics[width=0.4 \textwidth]{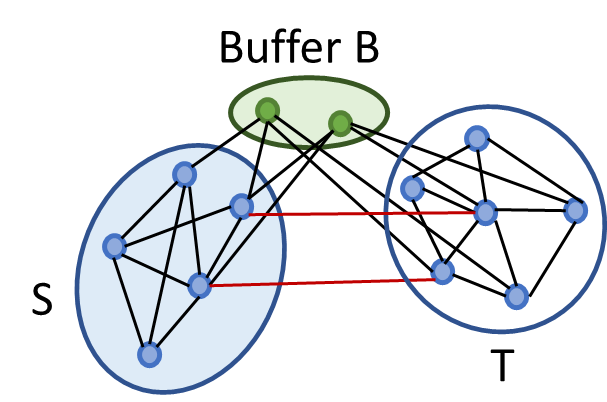} }}%
    \qquad
    \subfloat[\centering Buffered partitioning $(P_1, \dots, P_4 \| B_1, \dots, B_4)$]{{\includegraphics[width=0.5 \textwidth]{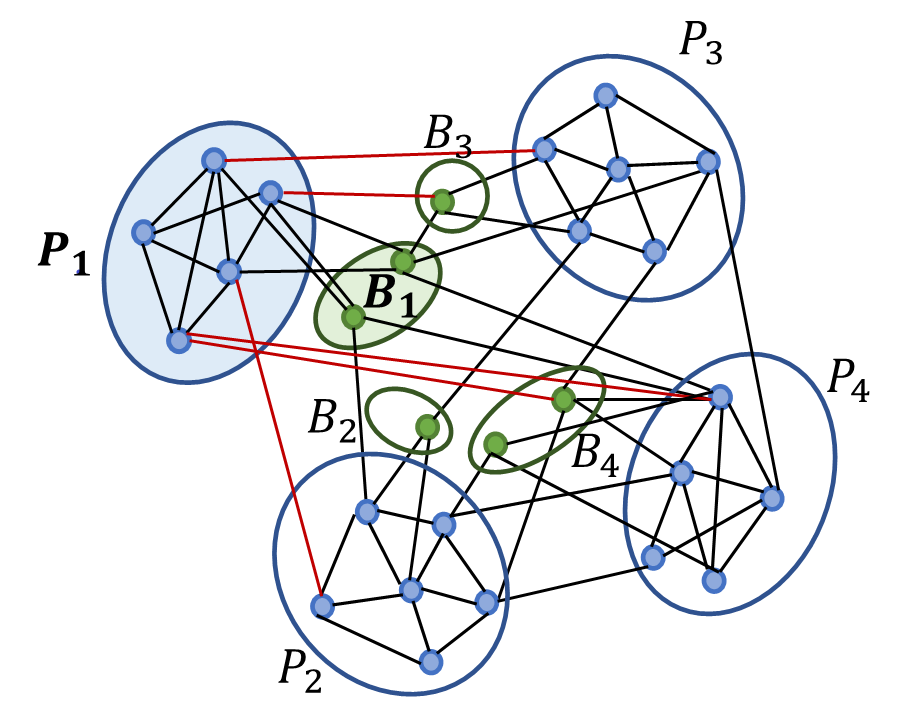} }}%
    \caption{{\em Left:} The figure on the left shows a partition of the vertex set $V$ into three pieces $S,B$ and $T=V \setminus (S \cup B)$. Here $B$ denotes the buffer, and cost of the this cut is $\delta(S,T)$, as denoted by the edges marked in red. The edges marked in grey denote the edges between $S$ and the buffer $B$. 
    \\
    {\em Right:} The illustrative figure shows a $k=4$ partition $P_1, P_2, P_3, P_4$ with buffers $B_1, B_2, B_3, B_4$. The {\em red} edges indicate the edges $\delta(P_1,V\setminus (P_1 \cup B_1))$ that contribute to the cut corresponding to $P_1$. All parts $P_1, \dots, P_4$ and $B_1,\dots, B_4$ are disjoint.}%
    \label{fig:buffer}%
\end{figure}

\newpage

\thispagestyle{empty}
\end{document}